\documentclass[11pt]{article}
\usepackage{amsmath,amsfonts,amssymb}
\usepackage{xcolor,graphicx}
\usepackage{mathrsfs,bbm}
\usepackage[numbers]{natbib}
\usepackage{geometry}
\usepackage[bookmarks=true, colorlinks, linkcolor=blue, urlcolor=blue, citecolor=blue, plainpages=false, pdfpagelabels, final, breaklinks=true]{hyperref}
\begin{document}

%%%% macros
\newcommand{\ii}{\operatorname{i}}
\newcommand{\EE}{\operatorname{EE}}
\newcommand{\LN}{\operatorname{LN}}
\newcommand{\drv}[1]{\frac{\partial}{\partial{#1}}}
\newcommand{\D}[1]{\mathop{\mathrm{D}#1}}
\newcommand{\Tr}{\operatorname*{Tr}}
\newcommand{\Pf}{\operatorname{Pf}}
\newcommand{\re}{\operatorname{Re}}
\newcommand{\im}{\operatorname{Im}}
\newcommand\BfC{\mathbb{C}}
\newcommand\hata{\hat{a}}
\newcommand\hatb{\hat{b}}
\newcommand\hatc{\hat{c}}
\newcommand\hatd{\hat{d}}
\newcommand\hatn{\hat{n}}
\newcommand\hatA{\hat{A}}
\newcommand\hatB{\hat{B}}
\newcommand\hatC{\hat{C}}
\newcommand\hatF{\hat{F}}
\newcommand\hatG{\hat{G}}
\newcommand\hatH{\hat{H}}
\newcommand\hatK{\hat{K}}
\newcommand\hatI{\hat{I}}
\newcommand\hatL{\hat{L}}
\newcommand\hatU{\hat{U}}
\newcommand\hatW{\hat{W}}
\newcommand\calC{\mathcal{C}}
\newcommand\calD{\mathcal{D}}
\newcommand\calG{\mathcal{G}}
\newcommand\varD{\mathscr{D}}
\newcommand\varF{\mathscr{F}}
\newcommand\varI{\mathscr{I}}
\newcommand\varM{\mathscr{M}}
\newcommand\varS{\mathscr{S}}
\newcommand\ket[1]{\ensuremath{\mathinner{\lvert{\textstyle#1}\rangle}}}
\newcommand\bra[1]{\ensuremath{\mathinner{\langle{\textstyle#1}\rvert}}}
\newcommand\avg[1]{\langle{\textstyle#1}\rangle}

\title{Efficient Gaussian Simulations of Fermionic Open Quantum Systems}
\author{Yinan Fang, Hyesung Choi, Minchul Lee, and Mahn-Soo Choi\thanks{choims@korea.ac.kr}}
\date{\small
School of Physics and Astronomy, Yunnan University, Yunnan 650500, China \\
Department of Physics, Korea University, Seoul 02841, Republic of Korea \\
School of Quantum, Korea University, Seoul 02841, Republic of Korea \\
Department of Applied Physics and Institute of Natural Science, College of Applied Science, Kyung Hee University, Yongin 17104, Korea \\
\ \\ \normalcolor
Revised December 19, 2025; Original September 14, 2025}
\maketitle

\begin{abstract}
We review existing classical simulation methods for performing fermionic Gaussian operations and develop new methods to address the gap by adhering to the fundamental theoretical framework established by Bravyi [Quantum Info. Comput. 5, 216 (2005)] for the most general fermionic Gaussian processes. Throughout this attempt, the focus remains on the unified approach that can be applied to generic fermionic Gaussian operations. This is beneficial since the selection of simulation methods has often been based on an ad hoc choice, heavily influenced by the specific model and circumstances, rather than on a systematic approach.
\end{abstract}

% \tableofcontents

\section{Introduction}

It is fermions that determine the properties of quantum materials, but the standard model of quantum computation is based on qubits.
The mapping between the Fock space of fermionic modes and traditional Hilbert space used for qubits enables simulation of fermionic systems with quantum circuits. However, this typically introduces overhead in ordinary qubit computers that must imitate fermionic statistics.
Therefore, it is beneficial to develop a quantum computation model \cite{Bravyi02a,Gonzalez-Cuadra23a} where algorithms and simulations run natively on registers of fermionic modes. Additionally, it is useful to analyze how algorithms can be designed to map fermionic gates to qubit gates on the standard model of quantum computation of qubits.

We note that recent advances have proposed and implemented platforms using free or trapped fermionic atoms (such as neutral atom arrays) in optical tweezers, which allow direct encoding of fermionic quantum information and realization of specialized fermionic quantum gates \cite{Gonzalez-Cuadra23a,2411.08955}.
These gates include tunneling (particle movement between sites) and interaction gates (such as occupation-based interactions), which mirror processes in many-body fermionic physics and quantum chemistry.
Fault-tolerant schemes, including repetition codes and color codes adapted to fermionic statistics, are under development for robust and scalable computation as well \cite{2411.08955,Li18d}.

On the other hand, a certain class of fermionic quantum operations, called fermionic Gaussian operations, can be efficiently simulated on classical computers \cite{Terhal02a,Bravyi05a,Bravyi12a,Oszmaniec14a,2411.18517}. 
Such fermionic Gaussian operations, such as the evolution by quadratic fermionic Hamiltonians (non-interacting fermion systems), can be efficiently simulated on classical computers using polynomial-time algorithms without exponential overhead. This makes them a well-understood and tractable subclass of fermionic quantum computations.
The ability to simulate these processes classically provides a benchmark and baseline for understanding when and how quantum advantage can be achieved by adding more complex non-Gaussian elements to the computation \cite{Mele25a,Dias24a}.
These classical simulations help identify the boundary between what quantum systems are simulable classically and what computations require genuine quantum resources. For example, fermionic Gaussian processes alone are not universal for quantum computing—they lack the “magic” needed for quantum advantage—so classical simulation shows their limitations and strengths \cite{Mele25a,2504.19317}.
Understanding Gaussian simulations clarifies the role of fermionic parity, Gaussianity, and superselection rules in the computational power and complexity of quantum circuits \cite{2411.18517}.

From practical point of view, classical algorithms for fermionic Gaussian simulations may provide powerful tools in quantum chemistry, condensed matter physics, and materials science where many physical fermionic systems can be approximated or decomposed into Gaussian components \cite{Mele25a,2504.19317}.
They enable efficient verification and benchmarking of quantum hardware by allowing classical checking of Gaussian fermionic operations implemented on quantum devices \cite{Dias24a}.
In particular, classical simulation of fermionic Gaussian processes has enabled the studies of measurement-induced entanglement transitions, which would be otherwise limited to only small systems and hinder clear understanding of the underlying physics \cite{2411.05671,Alberton21a,Cao19a,Piccitto23a,Li18c}.

This article combines a brief review with seminal results: 
it reviews some existing classical simulation methods for fermionic Gaussian operations and develops new methods to fill the gap by adopting the fundamental theoretical ground laid by Bravyi \cite{Bravyi05a} for the most general fermionic Gaussian processes.
Specifically, we propose a method to handle the non-unitary time evolution governed by an arbitrary Bogoliubov-de Gennes-type non-Hermitian Hamiltonian with the pairing potential. We also generalize the fermion measurement and dissipation of bare modes to deal with arbitrary dressed modes. Furthermore, we extend the decoherence process to include the projective jump operators associated with arbitrary dressed modes as well as the fermion dissipation operators. Finally, we provide simulation methods of random fermionic quantum circuits that can handle unitary gates involving the entire fermion modes (not just a pair of qubits in a brick wall-type pattern), fermion measurements, and fermion dissipation.
Throughout this attempt, we focus on the unified approach that can be applied to the generic fermionic Gaussian operations. This is instrumental since the selection of simulation methods appears to have been based on an ad hoc choice, strongly influenced by the specific model and circumstances rather than a systematic approach.

A strong motivation for writing this half-review is the breadth of systems and areas where classical simulations of fermionic Gaussian operations are used. The hope is that this work will assist in stimulating discussions and information transfer between different domains.
There are already examples that demonstrate the benefits of such an interaction. 
Recently, there have been some experimental efforts \cite{Roch14a,Mazzucchi16a,Koh23a} to observe or verify the effects of fermionic Gaussian processes predicted through classical simulations.
It is also interesting to note that there is an attempt to generalize classical simulation methods to non-Gaussian fermionic operations \cite{Dias24a}.

This article is organized as follows: 
Section~\ref{Paper::sec:Theory} provides an overview of the basic principles of fermionic Gaussian states and Gaussian superoperators (Section~\ref{Paper::sec:Principles}) and introduces or develops the classical simulation methods for various physical situations including non-Hermitian Hamiltonian dynamics (Section~\ref{Paper::sec:HamiltonianDynamics}), fermion measurement (Section~\ref{Paper::sec:FermiMeasurement}), fermion dissipation (Section~\ref{Paper::sec:FermiDissipation}), fermionic random quantum circuits (Section~\ref{Paper::sec:RandomCircuits}), quantum master equations (Section~\ref{Paper::sec:QME}) and continuous monitoring of fermion modes (Section~\ref{Paper::sec:ContMonitor}).
Section~\ref{Paper::sec:Models} takes two physical models, the Kitaev chain of Majorana fermions (Section~\ref{Paper::sec:KitaevChain}) and the Hatano-Nelson model for the non-Hermitian skin effects (Section~\ref{Paper::sec:HatanoNelson}), to demonstrate the classical simulation methods discussed in Section~\ref{Paper::sec:Theory}.

\section{Theory}
\label{Paper::sec:Theory}

A polynomial time simulation method for fermionic gates of the Gaussian type was first proposed by Terhal and DiVicenzo \cite{Terhal02a} motivated by a set of special quantum gates called ``matchgates'' \cite{Valiant02a,Knill01c} that allow for an efficient classical simulation.
Soon after, the theory of fermionic quantum computation was developed \cite{Bravyi02a}.
Since then, many variants of Terhal and DiVincenzo's method have been suggested for classical simulation of fermionic Gaussian processes.
Among them, Bravyi \cite{Bravyi05a} provides the most general theoretical ground.
Here, we follow the formulation in \cite{Bravyi05a} and suggest the unified classical simulation methods for fermionic Gaussian processes.

\subsection{Basic Principles}
\label{Paper::sec:Principles}

Fermions are typically described by Dirac fermion operators $\hata_i$ and $\hata_i^\dag$, which satisfy the anti-commutation relations
\begin{equation}
\label{Paper::eq:cmm:Dirac}
\{\hata_i,\hata_j^\dag\} = \delta_{ij} \,,\quad
\{\hata_i,\hata_j\} = \{\hata_i^\dag,\hata_j^\dag\} = 0.
\end{equation}
Each subscript index designates a collection of the spatial modes and internal degrees of freedom (e.g., spin) of the fermions.
In many cases, it is convenient to rewrite each Dirac fermion operator $\hata_i$ as a combination of two Majorana fermion operators
\begin{equation}
\label{Paper::eq:DiracMajorana}
\hata_k := \frac{1}{2}\left(\hatc_{2k-1} + \ii\hatc_{2k}\right).
\end{equation}
Majorana fermion operators satisfy the anti-commutation relations
\begin{equation}
\label{Paper::eq:cmm:Majorana}
\{\hatc_i,\hatc_j\} = 2\delta_{ij}
\end{equation}
in accordance with Eqs.~\eqref{Paper::eq:cmm:Dirac} and \eqref{Paper::eq:DiracMajorana};
in particular, $\hatc_i^2=1$ for any mode $i$.
More importantly, they are Hermitian, 
\begin{math}
\hatc_k^\dag = \hatc_k,
\end{math}
reflecting that Majorana fermions are anti-particles of their own.
Mathematically, both descriptions in Dirac and Majorana fermion operators are completely equivalent.
In this article, we will mainly use the Majorana fermion operators (rather than the Dirac fermion operators) since the former are easier to handle in numerical simulations as we will see below.

A many-fermion state is said to be Gaussian if it does not contain many-body correlation. More precisely, a fermionic state is a Gaussian state if and only if the Wick theorem \cite{Mahan00a,Zinn-Justin02a} holds. For example, the fourth-order moment of fermion operators may be expressed in terms of the second-order moment:
\begin{equation}
\avg{\hatc_1\hatc_2\hatc_3\hatc_4}
= \avg{\hatc_1\hatc_2}\avg{\hatc_3\hatc_4}
- \avg{\hatc_1\hatc_3}\avg{\hatc_2\hatc_4}
- \avg{\hatc_1\hatc_4}\avg{\hatc_2\hatc_3}.
\end{equation}
This generalizes to any higher-order correlation functions; see Eq.~\eqref{eq:higherCorr}.
Any $N$-fermion Gaussian pure state has the form
\begin{equation}
\label{Paper::eq:GaussianPure}
\ket{\Psi}_G = \hatb_1^\dag\hatb_2^\dag\cdots\hatb_N^\dag\ket{\ }_b,
\end{equation}
where $\hatb_k$ are ``dressed" fermion modes that are related to the bare fermion modes $\hata_k$ by the Bogoliubov-de Gennes (BdG) transformation
\begin{equation}
\label{Paper::eq:BdG}
\hatb_i 
= \sum_j\left(U_{ij}\hata_i + V_{ij}\hata_i^\dag\right)
\end{equation}
and $\ket{\ }_b$ is the vacuum state of the dressed modes $\hatb_k$, $\hatb_k\ket{\ }_b=0$. Note that the BdG transformation is canonical, i.e., preserves the anti-commutation relations so that
\begin{equation}
\label{Paper::eq:cmm:Dirac2}
\{\hatb_i,\hatb_j^\dag\} = \delta_{ij} \,,\quad
\{\hatb_i,\hatb_j\} = \{\hatb_i^\dag,\hatb_j^\dag\} = 0.
\end{equation}
This means that the subblock matrices $U$ and $V$ in Eq.~\eqref{Paper::eq:BdG} satisfy the unitarity relation
\begin{equation}
\begin{bmatrix}
U & V \\
V^* & U^*
\end{bmatrix}
\begin{bmatrix}
U^\dag & V^T \\
V^\dag & U^T
\end{bmatrix} =
\begin{bmatrix}
I & 0 \\
0 & I
\end{bmatrix}.
\end{equation}
The earlier simulation method \cite{Terhal02a} relied on the form \eqref{Paper::eq:GaussianPure} of the Gaussian state. Many classical simulations for the number-conserving processes, where $V=0$, are still using the form as well \cite{Feng24a,Piccitto23a,Kawabata23a,Russomanno23a,Cao19a}. However, they cannot describe the Gaussian mixed states, which is crucial for quantum noisy channels. More importantly, these methods break down or get highly inefficient for number-non-conserving processes ($V\neq 0$) and/or only apply to some restricted situations (such as continuous monitoring).

In this article, for the universal treatment of the Gaussian pure and mixed states and for efficient simulation of a wider range of systems especially with pairing correlations, which do not preserve the total number of particles and hence 
$V\neq 0$,
we adopt the form based on the Grassmann representation put forward by Bravyi \cite{Bravyi05a}.
A set of $2n$ Majorana fermion operators
\begin{math}
\{\hatc_1,\hatc_2,\cdots,\hatc_{2n}\},
\end{math}
satisfying the anti-commutation relations in Eq.~\eqref{Paper::eq:cmm:Majorana},
generates the Clifford algebra $\calC_{2n}$ consisting of polynomials of the generators $\hatc_k$.
On the other hand, the Grassmann algebra $\calG_{2n}$ is generated by an equal number of Grassmann variables
\begin{math}
\{x_1,x_2,\cdots,x_{2n}\}
\end{math}
satisfying the anti-commutation relations in Eq.~\eqref{Paper::eq:cmm:Grassmann}, and comprises polynomials of the generators $x_k$; see Appendix~\ref{Paper::sec:Grassmann}.
Now, consider a vector-space isomorphism between $\calC_{2n}$ and $\calG_{2n}$ defined by the correspondences
\begin{equation}
\label{Paper::eq:CliffordGrassmann}
\hatc_{i_1}\hatc_{i_2}\cdots\hatc_{i_k} \mapsto
x_{i_1}x_{i_2}\cdots x_{i_k} \,,\quad
\hatI \mapsto 1.
\end{equation}
Note that this isomorphism applies to arbitrary polynomials in $\calC_{2n}$ by the implied linearity. 
For a linear operator (i.e., polynomial in $\hatc_k$) $\hatA$ in $\calC_{2n}$, the image of $\hatA$ is called the \emph{Grassmann representation} of $\hatA$ and is denoted by $\hatA[x]$.
It is important to note that the above linear map in Eq.~\eqref{Paper::eq:CliffordGrassmann} is an isomorphism between $\calC_{2n}$ and $\calG_{2n}$ as vector spaces but does not preserve the multiplication in $\calC_{2n}$ and $\calG_{2n}$ as algebras. That is, $\hatA[x]\hatB[x]\neq (\hatA\hatB)[x]$ for two linear operators $\hatA$ and $\hatB$ in $\calC_{2n}$; see Eq.~\eqref{Paper::eq:superoperator} to get the Grassmann representation of $\hatA\hatB$ for relevant cases.

Bravyi \cite{Bravyi05a} provided the most general form of the Gaussian state by showing that a quantum state $\hat\rho$ of $2n$ Majorana fermion modes ($n$ Dirac fermion modes) is Gaussian if and only if $\hat\rho$ has a Grassmann representation of the form
\begin{equation}
\label{Paper::eq:GaussianState}
\hat\rho[x] =
\frac{P}{2^n}\exp\left(\frac{\ii}{2}\sum_{i,j=1}^{2n}x_iV_{ij}x_j\right)
\end{equation}
for some $2n\times 2n$ real anti-symmetric matrix $V$ and some constant $P$.
Here are a few remarks:
First, $\Tr\hat\rho=P$ and hence $P$ must be unity for a properly normalized state. For convenience, however, we leave $P$ as a parameter in order to handle trace non-conserving processes (such as measurements) in a universal way. 
It also provides an efficient way when splitting a numerical simulation into several intermediate steps.
Second, $\hat\rho$ itself cannot be written as an exponential function of an
operator in general because it may not have a full rank.
A notable example is 
\begin{equation}
\hat\rho = \hata\hata^\dag = \frac{1}{2} \left(\hatI -\ii\hatc_1\hatc_2\right),
\end{equation}
where $\hata:=(\hatc_1+\ii\hatc_2)/2$,
which describes the vacuum state $\ket{0}$ of $\hat{a}$, i.e., a pure Gaussian number state.
While it does not allow for an exponential form in the operator algebra $\calC_{2}$, its Grassmann representation takes the exponential form
\begin{equation}
\hat\rho[x] = \frac{1}{2}\exp\left(-\ii x_1x_2\right),
\end{equation}
which is consistent with Eq.~\eqref{Paper::eq:GaussianState}.
Third, by definition, all Gaussian states are even polynomials in the Grassmann representation.

The real anti-symmetric matrix $V$ in Eq.~\eqref{Paper::eq:GaussianState}, which we call the \emph{covariance matrix} of quantum state $\hat\rho$, contains the full information of $\hat\rho$. More precisely, due to the Wick theorem,
any higher-order correlation function of Majorana fermion operators is given in terms of covariance matrix $V$ by
\begin{equation}
\label{eq:higherCorr}
\avg{\hatc_{i_1}\hatc_{i_2}\cdots\hatc_{i_{2k}}}
:= \Tr\left[\hat\rho\,\hatc_{i_1}\hatc_{i_2}\cdots\hatc_{i_{2k}}\right]
= \ii^{-k}\Pf\left(\left.V\right|_{i_1i_2\cdots i_{2k}}\right) ,
\end{equation}
where
\begin{math}
\left.V\right|_{i_1i_2\cdots i_{2k}}
\end{math}
is the $2k\times 2k$ submatrix of $V$ with indicated rows and columns and $\Pf$ denotes the Pfaffian of an anti-symmetric matrix. In particular, the correlation matrix $C$ with elements $C_{ij}:=\avg{\hatc_i\hatc_j}$ is directly related to the covariance matrix $V$ by 
\begin{equation}
\label{eq:C-V}
C = I - \ii V.
\end{equation}
Therefore, strictly speaking, for a Gaussian operator $\hat\rho$ to describe a physical state (density operator), it must hold that $0\leq C\leq 1$ or equivalent equivalently $V^TV\leq 1$. 
In particular, a Gaussian state $\hat\rho$ is pure if and only if its covariant matrix $V$ is orthogonal, $VV^T=I$, or equivalently $V^2=-I$ since $V$ is anti-symmetric.
The relation~\eqref{eq:C-V} between the covariance matrix $V$ and the correlation matrix $C$ implies that for Gaussian states, quantum entanglement quantities such as logarithmic negativity, which is usually difficult to calculate for large systems, can be efficiently calculated from the covariance matrix; see Refs.~\cite{Shapourian17a,Shapourian19a,Shapourian19b}.

A linear map $\varF$ from $\calC_{2n}$ to $\calC_{2n}$, which we call a \emph{superoperator} throughout this article, is Gaussian if and only if for all linear operators $\hat\rho$ in $\calC_{2n}$,
\begin{equation}
\label{Paper::eq:superoperator}
\varF(\hat\rho)[x] = f_\varF\int\varD[y]\varD[z]\,
\exp{\left[ \ii 
\varS_\varF(x,y)
+ \ii y^T z \right]}\hat\rho[z],
\end{equation}
where $f_\varF$ is a complex prefactor and
\begin{equation}
\label{Paper::eq:action}
\varS_\varF(x,y) := \frac{1}{2}
\begin{bmatrix}
x^T & y^T
\end{bmatrix}
\begin{bmatrix}
A & B \\ 
-B^T & D
\end{bmatrix}
\begin{bmatrix}
x \\ y
\end{bmatrix}
\end{equation}
is the action of $\varF$ with some $2n\times 2n$ complex matrices $A$, $B$ and $D$ \cite{Bravyi05a}.
This Grassmann integral representation of a Gaussian superoperator $\varF$ provides a powerful tool to efficiently describe fermionic Gaussian processes as the covariance matrix $V_\varF$ and normalization factor $P_\varF:=\Tr[\varF(\hat\rho)]$ of $\varF(\hat\rho)$ are directly given in terms of matrices $A$, $B$ and $D$ in the action \eqref{Paper::eq:action} by
\begin{equation}
\label{Paper::eq:covariance}
V_\varF = A + B(I + VD)^{-1}VB^T
\end{equation}
and
\begin{equation}
\label{Paper::eq:Pfaffian:P}
P_\varF = (-1)^{n}f_\varF\Pf(V)\Pf(V^{-1}+D)\,P,
\end{equation}
respectively \cite{Bravyi05a,endnote:1}. 
When only the absolute value of $P_\varF$ is required, which is the case in many situations, the Pfaffian calculation may be avoided for the numerically more efficient determinant calculation
\begin{equation}
\label{Paper::eq:Det:P2}
P_\varF^2 = f_\varF^2\det(I + DV)P^2.
\end{equation}
An interesting consequence is that $\varF$ is trace-preserving if and only if $D=0$ and $f_\varF=1$.

A superoperator $\varF:\calC_{2n}\to\calC_{2n}$ is completely characterized by the corresponding Choi operator $\hatC_\varF$ in $\calC_{2n}\otimes\calC_{2n}$, known as the Choi-Jamiolkowski (CJ) isomorphism \cite{Jamiolkowski72a,Choi75a}.
Bravyi \cite{Bravyi05a} argued that for fermionic systems, it is more convenient to consider an alternative isomorphism between $\varF:\calC_{2n}\to\calC_{2n}$ and $\hat\rho_\varF$ in $\calC_{2n}\otimes_F\calC_{2n}$. Here, the fermionic tensor-product $\otimes_F$ is defined by
\begin{equation}
\calC_{2n}\otimes_F\calC_{2n} := \operatorname{Gen}
\{\hatc_1,\cdots,\hatc_{2n},\hatd_1,\cdots,\hatd_{2n}\}
\simeq \calC_{4n},
\end{equation}
where the generators $\hatc_k$ for the second Clifford algebra have been renamed $\hatd_k$ and anti-commutation $\{\hatc_i,\hatd_j\}=0$ 
(as well as $\{d_i,d_j\}=2\delta_{ij}$)
is implied.
Concretely, under this fermionic version of CJ isomorphism,
superoperator $\varF$ corresponds to the dual operator
\begin{equation}
\label{Paper::eq:CJB}
\hat\rho_\varF := (\varF\otimes_F\varI)\hat\rho_\Phi,
\end{equation}
where 
\begin{equation}
\label{Paper::eq:Bell}
\hat\rho_\Phi := \frac{1}{2^{2n}}
\prod_{i=1}^{2n}\left(I + \ii\hatc_i\hatd_i\right)
\end{equation}
is a Gaussian pure state in $\calC_{2n}\otimes_F\calC_{2n}$.
In Eq.~\eqref{Paper::eq:CJB}, $\varF\otimes_F\varI$ is a fermionic version of the tensor product between $\varF$ and identity superoperator $\varI$, defined by 
\begin{equation}
(\varF\otimes_F\varI)\left(\hatc_{i_1}\cdots\hatc_{i_k}
\hatd_{j_1}\cdots\hatd_{j_l}\right)
= \varF(\hatc_{i_1}\cdots\hatc_{i_k})\hatd_{j_1}\cdots\hatd_{j_l}
\end{equation}
with linearity implied.
For a Gaussian superoperator $\varF$ with the Grassmann integral representation in Eq.~\eqref{Paper::eq:superoperator}, the corresponding dual operator in Eq.~\eqref{Paper::eq:CJB} has the Grassmann representation
\begin{equation}
\label{Paper::eq:dualOp}
\hat\rho_\varF[x,y] 
= \frac{f_\varF}{2^{2n}}\exp\left[\ii\varS_F(x,y)\right].
\end{equation}
This means that the covariance matrix of $\hat\rho_\varF$ is given by
\begin{equation}
\label{Paper::eq:dual:covariance}
V_\varF := 
\begin{bmatrix}
A & B \\
-B^T & D
\end{bmatrix}.
\end{equation}
Then, it is clear that a Gaussian superoperator $\varF$ is \emph{completely positive} if and only if $V_\varF^TV_\varF\leq 1$ and $f_\varF>0$ \cite{Bravyi05a}.
Besides this important theorem, the simple relation~\eqref{Paper::eq:dualOp} between a Gaussian superoperators and its dual operator gives another powerful tool to describe fermionic Gaussian processes.

Any physically meaningful superoperator $\varF$ must be a completely positive linear map \cite{Breuer02a,ChoiMS22a}. This allows for the Kraus operator-sum representation of the form
\begin{equation}
\label{eq:KrausRep}
\varF(\hat\rho) = \sum_\mu \hat{F}_\mu \hat\rho\hat{F}_\mu^\dagger,
\end{equation}
where $\hat{F}_\mu$ are the Kraus operators describing specific decoherence processes.
In the following subsections, we discuss examples of Kraus operators that preserve the Gaussianity and are commonly encountered in practical physical situations.
For benchmark tests and scientific applications, we provide a publicly available implementation of all these elements through the symbolic quantum simulation framework \textsf{Q3} \cite{Q3}; see also Ref.~\cite{ChoiMS22a}.

\subsection{Hermitian and Non-Hermitian Hamiltonian Dynamics}
\label{Paper::sec:HamiltonianDynamics}

Consider the dynamics of quantum state
\begin{equation}
\label{Paper::eq:HamiltonianDynamics}
\hat\rho(t) = \hatW(t) \hat\rho(0)\hatW^\dag(t)
\end{equation}
governed through the time-evolution operator
\begin{equation}
\hatW(t) := \exp(-\ii t\hatK),
\end{equation}
by a quadratic Hamiltonian $\hatK$,
\begin{equation}
\hatK = \frac{\ii}{4}\sum_{i,j}\hatc_iK_{ij}\hatc_j + \text{(constant)}
\end{equation}
for an anti-symmetric matrix $K$.
When matrix $K$ is real, Hamiltonian $\hatK$ is Hermitian and the time-evolution operator $\hatW(t)$ is unitary describing the coherent dynamics of a closed system.
When $K$ is complex, on the other hand, $\hatK$ is non-Hermitian and $\hatW(t)$ is non-unitary. Such a non-Hermitian Hamiltonian dynamics has been extensively studied to phenomenologically describe the quantum decoherence process of open quantum systems or to analyze the post-selection data.

The dynamics described in Eq.~\eqref{Paper::eq:HamiltonianDynamics} is deterministic in the sense that the quantum state is determined once the initial state is fixed. Taking the derivative with respect to time $t$ leads to an equation of motion for $\hat\rho(t)$
\begin{equation}
\label{eq:nonHermitianDyn}
\frac{d\hat\rho}{dt} = -i\hatK\hat\rho + i\hat\rho\hatK^\dag.
\end{equation}
To make the coherent and incoherent dynamics clearer, 
we split the overall Hamiltonian $\hatK$ into the Hermitian and non-Hermitian parts
\begin{equation}
\hatK = \hatH -\ii \hat\Gamma
\end{equation}
with 
\begin{equation}
\label{eq:H:Gamma}
\hatH = \frac{\ii}{4}\sum_{i,j}\hatc_iH_{ij}\hatc_j \,,\quad
\hat\Gamma = \frac{\ii}{4}\sum_{i,j}\hatc_i\Gamma_{ij}\hatc_j + \Gamma_0,
\end{equation}
where $H$ and $\Gamma$ are real anti-symmetric matrices and $\Gamma_0$ is a positive parameter.

Since a fermionic Gaussian state $\rho(t)$ is completely characterized by covariance matrix $V(t)$ and prefactor $P(t)$, we rewrite the equation of motion~\eqref{eq:nonHermitianDyn} in terms of $V(t)$ and $P(t)$.
To do that, we put $H$ and $\Gamma$ in Eq.~\eqref{eq:H:Gamma}
and generalized Gaussian form~\eqref{Paper::eq:GaussianState} into Eq.~~\eqref{eq:nonHermitianDyn}, multiply the both sides with $\hatc_i\hatc_j$ and take the trace over the Hilbert space. Then, we get the equation of motion for the covariance matrix
\begin{equation}
\label{Paper::eq:HamitlonianDynamics2a}
\frac{dV}{dt} = [H,V] - (\Gamma + V\Gamma V) .
\end{equation}
Recall that the trace is not conserved for non-Hermitian Hamiltonian dynamics, and the normalization constant $P$ of $\hat\rho$ in \eqref{Paper::eq:GaussianState} is governed by
\begin{equation}
\label{Paper::eq:HamitlonianDynamics2b}
\frac{dP}{dt}
= -2P\left[\Gamma_0 - \frac{1}{4}\Tr\left(\Gamma V\right)\right],
\end{equation}
where $\Tr$ denotes the matrix trace.
By solving Eqs.~\eqref{Paper::eq:HamitlonianDynamics2a} and \eqref{Paper::eq:HamitlonianDynamics2b}, one can efficiently simulate any Hermitian or non-Hermitian Hamiltonian dynamics of the Gaussian type.
Equation~\eqref{Paper::eq:HamitlonianDynamics2a} is non-linear in the covariance matrix $V$ and does not allow for an analytic solution. However, we find that the fourth-order Runge-Kutta methods and other similar methods are enough for accurate and efficient numerical solution of the equation.

\subsection{Fermi Measurement}
\label{Paper::sec:FermiMeasurement}

Consider a measurement process of the occupation number
\begin{equation}
\hatn_i := \hata_i^\dag\hata_i
\end{equation}
for a \emph{bare} fermion mode $\hata_i$ \cite{Bravyi05a}.
The measurement outcome is either $n_i=0$ or $1$.
When the measurement outcome is 1, the state ``collapses'' as described by the superoperator
\begin{equation}
\label{Paper::eq:outcome:1}
\varM_{i,1}:\hat\rho \mapsto 
\hat\rho_{\varM_{i,1}} :=
\hatn_i\hat\rho\,\hatn_i.
\end{equation}
The probability for this process to actually occur is given by
\begin{equation}
p_1 = \Tr\left[\hatn_i\hat\rho\, \hatn_i\right]  
= \Tr\left[\hat\rho\,\hatn_i\right].
\end{equation}
It is convenient to first calculate the dual operator $\hat\rho_{\varM_{i,1}}$ and its Grassmann representation \cite{Bravyi05a} to get
\begin{equation}
\hat\rho_{\varM_{i,1}}[x,y] = \frac{1}{2^{2n}}
\exp\left[\ii\varS_{\varM_{i,1}}(x,y)\right],
\end{equation}
with
\begin{equation}
\varS_{\varM_{i,1}}(x,y) = \frac{1}{2}
\begin{bmatrix}
x^T & y^T
\end{bmatrix}
\begin{bmatrix}
A & B \\
-B^T & -A
\end{bmatrix}
\begin{bmatrix}
x \\ y
\end{bmatrix},
\end{equation}
where the matrix elements of $2n\times 2n$ matrices $A$ and $B$ are given by
\begin{subequations}
\begin{align}
A_{pq} &= \delta_{p,2i-1}\delta_{q,2i} - \delta_{p,2i}\delta_{q,2i-1} ,\\
B_{pq} &= \delta_{pq} - 
\delta_{p,2i-1}\delta_{q,2i-1} - \delta_{p,2i}\delta_{q,2i}.
\end{align}
\end{subequations}
Then, from the relation Eq.~\eqref{Paper::eq:dualOp} between the integral representation of a superoperator and the Grassmann representation of its dual operator, we can get the covariance matrix $V_{\varM_{i,1}}$ of the post-measurement state [see Eq.~\eqref{Paper::eq:covariance}]
\begin{equation}
\label{Paper::eq:measurement:V1}
V_{\varM_{i,1}} = A + B(I - VA)^{-1}VB^{T}
\end{equation}
and the prefactor [see Eq.~\eqref{Paper::eq:Pfaffian:P}]
\begin{equation}
\label{Paper::eq:measurement:P1}
P_{\varM_{i,1}} = \frac{1}{2}\sqrt{\det(I-AV)} .
\end{equation}
Similarly, for the measurement outcome $0$,
\begin{equation}
\label{Paper::eq:outcome:0}
\varM_{i,0}:\hat\rho \mapsto 
\hat\rho_{\varM_{i,0}} :=
\hata_i\hata_i^\dag\hat\rho\,\hata_i\hata_i^\dag,
\end{equation}
the covariance matrix and prefactor are given by 
\begin{equation}
\label{Paper::eq:measurement:V0}
V_{\varM_{i,0}} = -A + B(I + VA)^{-1}VB^{T}
\end{equation}
and
\begin{equation}
\label{Paper::eq:measurement:P0}
P_{\varM_{i,0}} = \frac{1}{2}\sqrt{\det(I+AV)} ,
\end{equation}
respectively.

We now generalize the measurement to the fermion number
\begin{equation}
\hatn_b := \hatb^\dag\hatb
\end{equation}
for an arbitrary (unnormalized) \emph{dressed} mode
\begin{equation}
\label{Paper::eq:dressedMode}
\hatb = \sum_{i=1}^{2n} w_i\hatc_i .
\end{equation}
Note that the dressed mode must satisfy the canonical anti-commutation relations
\begin{equation}
\label{Paper::eq:dressdedMode1a}
\{\hatb,\hatb^\dag\} = 1 \,,\quad
\{\hatb,\hatb\} = \{\hatb^\dag,\hatb^\dag\} = 0,
\end{equation}
which implies the conditions
\begin{equation}
\label{Paper::eq:dressdedMode1b}
u^Tv = 0 \,,\quad
\left\|u\right\| = \left\|v\right\| = \frac{1}{2}
\end{equation}
where $u:=\re{w}$, $v:=\im{w}$, and $w^T:=(w_1,w_2,\cdots,w_{2n})$.

To calculate $\varM_{b,1}$ and $\varM_{b,0}$ corresponding to the measurement outcome 0 and 1, respectively, one can follow the above method to get the dual operators.
Instead, here we will use a canonical transformation:
Let us choose an orthogonal set of dressed modes
\begin{math}
\{\hatb_1\equiv\hatb,\hatb_2,\cdots,\hatb_n\};
\end{math}
recall the anti-commutation relations
\begin{math}
\{\hatb_i,\hatb_j^\dag\} = \delta_{ij}
\end{math}
and
\begin{math}
\{\hatb_i,\hatb_j\} = \{\hatb_i^\dag,\hatb_j^\dag\} = 0.
\end{math}
In terms of the dressed Majorana fermion modes, 
\begin{math}
\hatd_{2i-1} := \hatb_i+\hatb_i^\dag
\end{math}
and
\begin{math}
\hatd_{2i} := -i(\hatb_i - \hatb_i^\dag),
\end{math}
the first two dressed Majorana fermion modes are related to the bare Majorana fermion modes by
\begin{equation}
\label{Paper::eq:dressed}
\hatd_1 := \sum_{i=1}^{2n} u_i\hatc_i \,,\quad
\hatd_2 := \sum_{i=1}^{2n} v_i\hatc_i .
\end{equation}
The idea is that if the Gaussian state is expressed in terms of these dressed modes $\hatd_j$ instead of the bare modes $\hatc_i$, 
then the measurement process can be described simply by $\varM_{i=1,0}$ [see Eq.~\eqref{Paper::eq:outcome:0}] and $\varM_{i=1,1}$ [see Eq.~\eqref{Paper::eq:outcome:1}] as dressed modes $\hatb_i$ play the role of bare modes $\hata_i$.
When the post-measurement state is required, the resulting state is transformed back to the form in terms of the bare modes.

Note that Eq.~\eqref{Paper::eq:dressed}, or equivalently $\hatb_1=\hatb$, is the only requirement and otherwise the choice of the orthogonal dressed modes is completely arbitrary.
This can be achieved by a $2n\times2n$ Householder reflection matrix $T$ such that for the $2\times2n$ matrix
\begin{equation}
M := 
\begin{bmatrix}
u_1 & u_2 & \cdots & u_{2n} \\
v_1 & v_2 & \cdots & v_{2n}
\end{bmatrix},
\end{equation}
$MT$ is a lower-triangular (in fact, diagonal since the vectors $u$ and $v$ are orthogonal to each other) matrix with every diagonal element positive \cite{Gallier01a}.
This transformation $T$ induces the canonical transformation from the bare modes $\hatc_i$ to dressed modes $\hatd_j$,
\begin{equation}
\label{Paper::eq:canonicalT}
\hatd_i = \sum_{j=1}^{2n}T^T_{ij}\hatc_j.
\end{equation}
Note that by manifestation, the first dressed Dirac fermion mode is nothing but the dressed mode being measured,
\begin{math}
\hatb_1 := (\hatd_1 +\ii\hatd_2)/2=\hatb.
\end{math}
Since $T$ is orthogonal, the dressed modes satisfy the canonical anti-commutation relations
\begin{math}
\{\hatd_i,\hatd_j\} = 2\delta_{ij}
\end{math}
as they should.
Since the covariance matrix $V$ transforms as $T^TVT$ under this canonical transformation, the superoperators $\varM_{b,m}$ for $m=1,0$ are simply given by $\varM_{1,m}$ [see Eqs.~\eqref{Paper::eq:measurement:V0}, \eqref{Paper::eq:measurement:P0}, \eqref{Paper::eq:measurement:V1}, and \eqref{Paper::eq:measurement:P1}] with the action subblock matrices $A$ and $B$ replaced by $TAT^T$ and $TBT^T$, respectively. 

If the Gaussian state is pure and the dressed mode $\hatb$ does not mix the particle and hole operators (i.e., either $w_{2i-1}=iw_{2i}$ or $w_{2i-1}=-iw_{2i}$), then one can use the QR decomposition method suggested by 
\cite{Feng24a,Piccitto23a}. This method, if applicable at all, is slightly faster since it does not involve the calculation of matrix inversion.

\subsection{Fermi Dissipation}
\label{Paper::sec:FermiDissipation}

We now consider the dissipation process
\begin{equation}
\varF(\hat\rho) = \hatb\hat\rho\,\hatb^\dag
\end{equation}
of a dressed Dirac fermion mode $\hatb$ as defined in Eq.~\eqref{Paper::eq:dressedMode} satisfying Eqs.~\eqref{Paper::eq:dressdedMode1a} and \eqref{Paper::eq:dressdedMode1b}.
In \cite{Bravyi05a}, the dissipated mode was one of the bare modes $\hata_i$. Here, we generalize the calculation to an arbitrary dressed mode.

It is convenient to first get the dual operator $\hat\rho_{\varF}$ in $\calC_{4n}$ corresponding to the superoperator $\varF$. Starting with Eqs.~\eqref{Paper::eq:CJB} and \eqref{Paper::eq:Bell}, after some algebraic calculations, one can get 
\begin{equation}
\hat\rho_{\varF_\mu}[x,y] = \frac{f_{\varF}}{2^{2n}}
\exp\left(
\frac{\ii}{2}
\begin{bmatrix}
x^T & y^T
\end{bmatrix}
\begin{bmatrix}
A & B \\
-B^T & A
\end{bmatrix}
\begin{bmatrix}
x\\ y
\end{bmatrix}
\right).
\end{equation}
The prefactor is given by $f_{\varF}=1/2$, and
the subblock matrices $A$ and $B$ are given by
\begin{subequations}
\label{Paper::eq:dissipative:AB}
\begin{align}
A &:= -4\left(u v^T - v u^T\right) \,,\\
B &:= I - 4\left(u u^T + v v^T\right) \,,
\end{align}
\end{subequations}
where vectors $u$ and $v$ are related to $w$ by Eq.~\eqref{Paper::eq:dressdedMode1b}.
By the fermionic version of the CJ isomorphism, they in turn characterize the integral representation of $\varF_\mu$ [see Eqs.~\eqref{Paper::eq:superoperator} and \eqref{Paper::eq:action}].
Plugging Eq.~\eqref{Paper::eq:dissipative:AB} into the formulae~\eqref{Paper::eq:covariance} and \eqref{Paper::eq:Pfaffian:P}, one can get the quantum state after the process has occurred.

As in the measurement process discussed in Section~\ref{Paper::sec:FermiMeasurement}, if the Gaussian state is pure and the dressed mode $\hatb$ does not mix the particle and hole operators (i.e., either $w_{2i-1}=iw_{2i}$ or $w_{2i-1}=-iw_{2i}$), then one can use the QR decomposition method suggested by \cite{Feng24a,Piccitto23a}.

\subsection{Fermionic Random Circuits}
\label{Paper::sec:RandomCircuits}

\begin{figure}
\centering
\includegraphics[width=100mm]{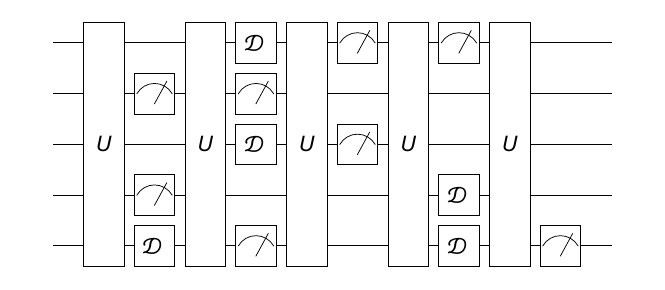}
\caption{A typical example of fermionic random circuit with alternating layers of unitary and measurement layers. Each unitary layer (labeled $U$) describes fermionic gate of the Gaussian form 
$\hatU:=\exp\left(\sum_{ij}\hatc_i H_{ij}\hatc_j\right)$,
where $\hatc_i$ are Majorana fermion operators and $H$ is a real anti-symmetric matrix. Each measurement layer consists of projective measurements (depicted as measurement devices) or dissipation processes (labeled $\calD$ in the picture) of randomly selected modes.}
\label{Paper::fig:1}
\end{figure}

We consider fermionic random circuits consisting of alternating layers of unitary and measurement layers as shown in Fig.~\ref{Paper::fig:1}.
Each unitary layer [labeled $U$ in Fig.~\ref{Paper::fig:1}] describes fermionic gate of the Gaussian form
\begin{equation}
\hatU := \exp\left(\sum_{ij}\hatc_i H_{ij}\hatc_j\right),
\end{equation}
where $H$ is a $2n\times 2n$ real anti-symmetric matrix.
Typically, the anti-symmetric matrix is randomly selected from a uniform distribution with respect to the Haar measure. In this work, we consider a few different distributions to reflect some physical situations.

The measurement layer consists of projective measurements [depicted as measurement devices in Fig.~\ref{Paper::fig:1}] of random dressed Dirac fermion modes $\hatn_\mu:=\hatb_\mu^\dag\hatb_\mu$
and the fermion dissipation [labeled $\calD$ in Fig.~\ref{Paper::fig:1}]
\begin{math}
\varF_\nu(\hat\rho) = \hatb_\nu\hat\rho\,\hatb_\nu^\dag .
\end{math}
Typically, $\hatb_\mu$ is randomly chosen among the bare modes $\hata_i$. In this work, we also consider some physically-motivated sets of dressed modes.
Measurements of dressed modes and fermion dissipation (of bare or dressed modes) would not allow a direct physical implementation in plausible fermionic quantum computers. However, it is known that in principle, these can be implemented by attaching ancillary fermion modes.

All the fermionic circuit elements mentioned above can be efficiently simulated using the methods in Sections~\ref{Paper::sec:HamiltonianDynamics} (for Gaussian unitary gates), \ref{Paper::sec:FermiMeasurement} (for projective measurements) and \ref{Paper::sec:FermiDissipation} (for fermion dissipation).

Note that most works on random quantum circuits consider two-qubit gates arranged in a brick-wall pattern (or some other structure parameterized quantum circuits). 
This is usually to avoid the high cost of numerical simulations for gate elements involving many fermion modes.
Since we have efficient simulation methods for Gaussian circuit elements the cost of which scales as a only low-degree polynomial, we will consider quantum circuits where the Gaussian unitary gate involves the whole fermion modes, rather than the brick-wall style circuits.

\subsection{Quantum Master Equation}
\label{Paper::sec:QME}

The Lindblad equation (also called quantum master equation) is the most general form of a Markovian master equation used to describe the non-unitary evolution of a quantum system interacting with its environment, ensuring completely positive and trace-preserving dynamics \cite{Breuer02a}.
The time evolution of the density matrix $\hat\rho$ for an open quantum system is governed by the Lindblad equation:
\begin{equation}
\label{Paper::eq:LindbladEq}
\frac{d\hat\rho}{dt} = -i\left[\hatH,\hat\rho\right]
+ \sum_\mu \gamma_\mu\left(\hatL_\mu\hat\rho\hatL_\mu^\dag
-\frac{1}{2}\hatL_\mu^\dag\hatL_\mu\hat\rho
-\frac{1}{2}\hat\rho\hatL_\mu^\dag\hatL_\mu
\right),
\end{equation}
where $\hatH$ is the system Hamiltonian governing the unitary evolution, $\hatL_\mu$ are the Lindblad operators modeling the decoherence processes and
$\gamma_\mu$ are the decoherence rates.
The main focus of our work is on the Lindblad equation for many-body systems of non-interacting fermions with Lindblad operators of either dissipative 
\begin{math}
(\hatL_\mu = \hatb_\mu)
\end{math}
or projective 
\begin{math}
(\hatL_\mu = \hatb_\mu^\dag\hatb_\mu)
\end{math}
type, where $\hatb_\mu$ are dressed Dirac fermion modes
\begin{equation}
\hatb_\mu := \sum_{i=1}^{2n}w_{\mu,i}\hatc_i.
\end{equation}
Recall that each $\hatb_\mu$ is a valid Dirac fermion operator (see Sections~\ref{Paper::sec:FermiMeasurement} and \ref{Paper::sec:FermiDissipation})
\begin{equation}
\{\hatb_\mu,\hatb_\mu^\dag\} = 1 \,,\quad
\{\hatb_\mu,\hatb_\mu\} = \{\hatb_\mu^\dag,\hatb_\mu^\dag\} = 0,
\end{equation}
which implies that [see also Eqs.~\eqref{Paper::eq:dressdedMode1a} and \eqref{Paper::eq:dressdedMode1b}]
\begin{equation}
u_{\mu}^Tv_{\mu} = 0 ,\quad
\left\|u_\mu\right\| = \left\|v_\mu\right\| = \frac{1}{2},
\end{equation}
where $u_{\mu}:=\re w_{\mu}$ and $v_{\mu}=\im w_{\mu}$, and 
\begin{math}
w_\mu^T := (w_{\mu,1},w_{\mu,2},\cdots,w_{\mu,2n}).
\end{math}
For distinct $\hatb_\mu$ and $\hatb_\nu$, however, there is no definite relation required between them; they refer to physically independent processes.

A typical approach is to solve (analytically or numerically) the Lindblad equation to get the density operator $\hat\rho(t)$.
This is useful when one merely wants to examine the expectation values of physical quantities. For numerical solutions, usually it requires a huge amount of memory and high computational costs when the system is large.
For the dissipative or projective case specified above, however, the Lindblad equation preserves the Gaussianity of quantum state 
[see Section~\ref{Paper::sec:Principles} and Eq.~\eqref{Paper::eq:GaussianState} in particular]
and can be solved efficiently.
Specifically, we multiply the both sides of Eq.~\eqref{Paper::eq:LindbladEq} by $\hatc_i\hatc_j$ and take the trace of terms to calculate various correlation functions by using the Wick theorem \cite{Mahan00a}.
In the dissipative case \cite{Bravyi12a},
this leads to the equation for the covariance matrix $V$
\begin{equation}
\label{Paper::eq:matV}
\frac{dV}{dt} 
= V(H-M-M^T)^T + (H-M-M^T)V - i2(M-M^T) \,,
\end{equation}
where matrix $M$ is defined by
\begin{math}
M := \sum_\mu \gamma_\mu M_\mu
\end{math}
with
\begin{math}
M_\mu := w_\mu w_\mu^\dag.
\end{math}
This equation is linear in $V$ and can be solved efficiently, e.g., by using the Bartels-Stewart algorithm \cite{Bravyi12a,Bartels72a}.

In the projective case, the corresponding equation is given by
\begin{multline}
\label{Paper::eq:matV2}
\frac{dV}{dt}
= 2\sum_\mu\gamma_\mu^2\Tr\left[(M_\mu-M_\mu^T)V\right]
\left(M_\mu - M_\mu^T\right) \\{}
+ \left[H-\sum_\mu\gamma_\mu^2(M_\mu+M_\mu^T)\right]V
+ V\left[H-\sum_\mu\gamma_\mu^2(M_\mu+M_\mu^T)\right]^T.
\end{multline}
This equation is non-linear in $V$ and requires higher computational costs. Nevertheless, numerical integration methods such as the fourth-order Lunge-Kutta methods may be applied to solve the equation in polynomial time.

Another powerful (and more interesting in our context) approach for simulating open quantum systems governed by the Lindblad equation is to unravel the deterministic Lindblad equation into an ensemble average over many individual pure-state quantum trajectories, each representing a possible evolution of the system subject to both continuous non-unitary evolution and random ``quantum jumps" generated by the dissipative environment \cite{Plenio98a}.
This so-called quantum trajectory approach (also known as the quantum jump approach or Monte Carlo wave-function method) is usually suggested to overcome the huge memory requirement for large systems.
However, the notion of quantum trajectories is crucial to analyze the dynamics of some features such as entropy and entanglement that are non-linear functions  of the state and cannot be attained from the density matrix (a sort of average of possible states). The trajectory approach enables to capture those non-linear features in the state evolving in time.
As a specific way to unravel the Lindblad equation, the stochastic Schr{\"o}dinger equation is commonly used \cite{Yip18a}. Here, we will follow \cite{Plenio98a} and discretize Eq.~\eqref{Paper::eq:LindbladEq} to describe the evolution from $\hat\rho(t)$ to $\hat\rho(t+dt)$ by the step-wise Kraus representation
\begin{equation}
\label{Paper::eq:discretizedQME}
\delta\varF(\hat\rho) \approx \hatF_0\hat\rho\hatF_0^\dag +
\sum_\mu\hatF_\mu\hat\rho\hatF_\mu^\dag ,
\end{equation}
where
\begin{equation}
\hatF_\mu := \sqrt{dt\gamma_\mu}\, \hatL_\mu
\end{equation}
and
\begin{equation}
\hatF_0 = \exp\left[-\ii dt\left(\hatH - \ii\hatG\right)\right]
\end{equation}
with
\begin{math}
\hatG := \frac{1}{2}\sum_\mu\gamma_\mu\hatL_\mu^\dag\hatL_\mu .
\end{math}
Within the discretization approximation, the superoperator preserves the trace:
\begin{equation}
\hatF_0^\dag\hatF_0 + \sum_\mu\hatF_\mu^\dag\hatF_\mu \approx I.
\end{equation}
Each term in Eq.~\eqref{Paper::eq:discretizedQME} refers to a possible random process to occur with probability
\begin{equation}
p_0 = \Tr\left[\hatF_0\hat\rho\hatF_0^\dag\right],\quad
p_\mu = \Tr\left[\hatF_\mu\hat\rho\hatF_\mu^\dag\right].
\end{equation}
When a particular process is realized, the state is reduced to
\begin{math}
p_0^{-1}\hatF_0\hat\rho\hatF_0^\dag
\end{math}
or
\begin{math}
p_\mu^{-1}\hatF_\mu\hat\rho\hatF_\mu^\dag.
\end{math}
The process $\hatF_0\hat\rho\hatF_0^\dag$ is a non-Hermitian Hamiltonian dynamics and can be evaluated by the method in Section~\ref{Paper::sec:HamiltonianDynamics}.
On the other hand, processes $\hatF_\mu\hat\rho\hatF_\mu^\dag$ can be simulated by the methods in Section~\ref{Paper::sec:FermiMeasurement} for the projective case and in Section~\ref{Paper::sec:FermiDissipation} for the dissipative case.
Repeating this step until the final time, a trajectory of quantum state is obtained. Repeating this whole procedure many times leads to an ensemble of quantum trajectories to perform statistical analysis with.
Here, the Kraus representation in Eq.~\eqref{Paper::eq:discretizedQME} arises as an approximate discretization of the Lindblad equation. Simulation methods of a genuine decoherence process described by the Kraus representation is discussed in Appendix~\ref{Paper::sec:NoisyChannels}.

Figure~\ref{Paper::fig:2} demonstrates how the simulation method works by comparing the two results of logarithmic entanglement negativity of the mixed-state density matrix from the direct numerical solution of the Lindblad equation 
[see Eqs.~\eqref{Paper::eq:matV} and \eqref{Paper::eq:matV2}]
and from our simulation method in the Hatano-Nelson model (to be discussed in detail in Section~\ref{Paper::sec:HatanoNelson}) for relatively small system ($L=6$) \cite{endnote:2}.

\begin{figure}
\centering
\includegraphics[width=70mm]{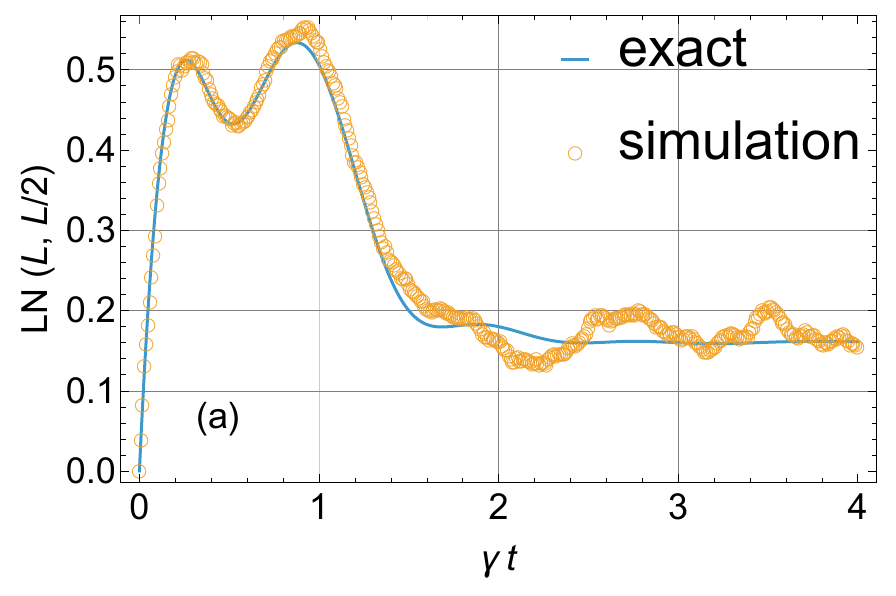}
\includegraphics[width=70mm]{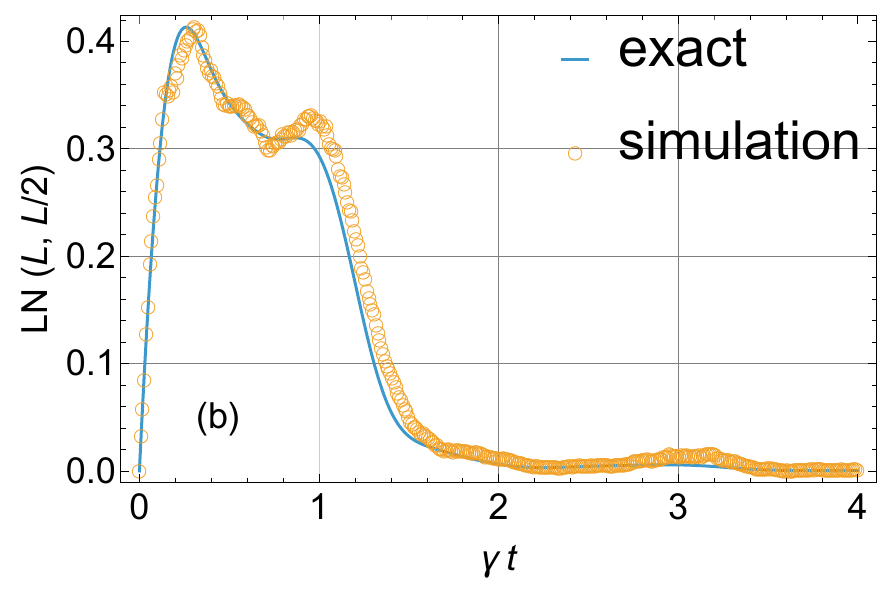}
\caption{Comparison of the logarithmic negativity $\LN(L,L/2)$ from direct numerical solution of the Lindblad equation (blue solid line) and Wick simulation (orange dots) in the dissipative (a) and projective (b) Hatano-Nelson model for $L=6$ and $\gamma/J=0.5$.
For the simulation, the step size in time was $dt=0.01/\gamma$ and the sample size was $N=500$.}
\label{Paper::fig:2}
\end{figure}

In the quantum trajectory approach, the quantum state in a trajectory may be regarded as a pure state. Consequently, as long as the dressed mode $\hatb$ does not mix the particle and hole operators (i.e., either $w_{\mu,2i-1}=iw_{\mu,2i}$ or $w_{\mu,2i-1}=-iw_{\mu,2i}$), one can use the QR decomposition method suggested by \cite{Feng24a,Piccitto23a}.

In the above, we have discussed the Markovian dynamics described by the Lindblad equation~\eqref{Paper::eq:LindbladEq}.
One can extend this method to non-Markovian cases associated with a time-nonlocal memory kernel by discretizing the time.

\subsection{Continuous Monitoring of Fermion Modes}
\label{Paper::sec:ContMonitor}

Another physically interesting situation of our concern is continuous monitoring of fermion modes.
Cao, Tilloy and De Luca \cite{Cao19a} devised an efficient simulation method for continuous monitoring of fermion modes in the number-conserving case. A slightly generalized form of this method was provided by Refs.~\cite{Piccitto23a,Russomanno23a,2411.05671} that can apply in the presence of pairing potential. 
These methods, however, are limited to continuous monitoring and cannot describe more general Gaussian quantum noise (such as described by linear or quadratic Lindblad operators).
Here, we describe a unified method based on the generic form~\eqref{Paper::eq:GaussianState} of the fermionic Gaussian state.

\begin{figure}
\centering
\includegraphics[width=90mm]{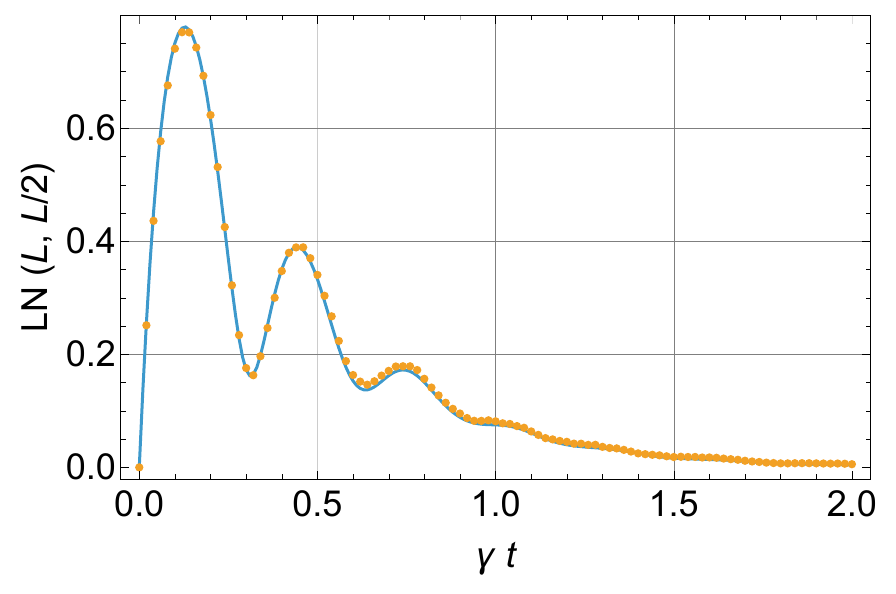}
\caption{Comparison of the mixed-state logarithmic negativity (LN) from the direct numerical solution of the Lindblad equation (blue solid line) and Monte Carlo simulation (orange dots) for the problem of continuous monitoring in the Kitaev chain of Majorana fermions.
In both cases, the chemical potential is $\mu/\Delta=0.2$, the hopping amplitude is $J/\Delta=1$, the monitoring rate is $\gamma/\Delta=0.1$ and the system size is $L=6$.
The simulation for $N=500$ quantum trajectories were performed with step size of time $dt=0.001/\gamma$, whereas  the values of logarithmic negativity are plotted in the step size $dt'=0.02/\gamma$.}
\label{Paper::fig:3}
\end{figure}

We model the situation in two different and closely related ways. 
First, we describe the continuous monitoring in terms of a Lindblad equation:
\begin{equation}
\label{Paper::eq:LindbladMonitor}
\frac{d\hat\rho}{dt} = -\ii\left[\hatH,\hat\rho\right] 
+ \sum_\mu\gamma_\mu\left(
	\hatb_\mu^\dag\hatb_\mu\hat\rho\,\hatb_\mu^\dag\hatb_\mu
	+ \hatb_\mu\hatb_\mu^\dag\hat\rho\,\hatb_\mu\hatb_\mu^\dag
	- \hat\rho
\right).
\end{equation} 
Here, all Lindblad operators are projection operators and appear in pairs,
\begin{math}
\hatb_\mu^\dag\hatb_\mu
\end{math}
and
\begin{math}
\hatb_\mu\hatb_\mu^\dag
\end{math}
corresponding to the opposite measurement outcomes, which add up to a constant.
It is thus an interesting special case of the Lindblad equation considered in Section~\ref{Paper::sec:QME}.
As a verification of the quantum trajectory approach, Fig.~\ref{Paper::fig:3} compares the logarithmic entanglement negativity of the mixed-state density matrix from the direct numerical solution of the Lindblad equation  
[see Eqs.~\eqref{Paper::eq:matV} and \eqref{Paper::eq:matV2}]
and the one from quantum trajectory simulation for the continuous monitoring in the Kitaev chain of Majorana fermions (to be discussed in detail in Section~\ref{Paper::sec:KitaevChain}) for a relatively small system ($L=6$) \cite{endnote:2}.

Second, we use an equivalent quantum circuit. To understand the underlying idea, note that discretizing the Lindblad equation leads to a quantum circuit model with unitary gates (with a decay factor, see below) and Fermi measurements as circuit elements; see Section~\ref{Paper::sec:RandomCircuits}. This is possible because Lindblad operators are projection operators and appear in pairs with opposite eigenvalues, i.e., $\hatb_\mu^\dag\hatb_\mu$ and $\hatb_\mu\hatb_\mu^\dag$ for dressed Dirac fermion modes $\hatb_\mu$. 
In fact, the same reason gives the motivation of the methods in \cite{Cao19a} and \cite{Piccitto23a,Russomanno23a,2411.05671} mentioned above.

\section{Physical Models}
\label{Paper::sec:Models}

In the previous section, we explored how to efficiently simulate various noisy fermionic quantum processes. 
In this section, we delve into several physical models and apply the methods discussed earlier. 
Our primary focus will be on showcasing the simulation methods rather than delving into specific physical effects that may be intriguing within the models. 
Anyone interested in these physical effects can apply the methods to conduct thorough investigations.

Among various quantum noisy dynamical features of fermionic open systems, 
we will mainly focus on the quantum entanglement. As mentioned before, since the quantum entanglement is a non-linear function of quantum states, the quantum trajectory approach captures its dynamics better than the density matrix. The difference will be illustrated (see Section~\ref{Paper::sec:HatanoNelson} in particular) by comparing the \emph{average entanglement}, i.e., the entanglement averaged over the trajectories of quantum states and the \emph{mixed-state entanglement}, i.e., the entanglement in the density matrix.

To quantify the quantum entanglement between different parts of a quantum system, various entanglement measures have been proposed \cite{Plenio07a}.
Common examples include entanglement entropy~\cite{Bennett96b}, concurrence~\cite{Wootters98a}, and logarithmic negativity (or negativity)~\cite{Plenio05a}. However, entanglement entropy is applicable only to pure states, while concurrence is practical only for two-qubit systems. 
For mixed states, logarithmic negativity is one of the rare practically computable entanglement measures. Nevertheless, its computational cost increases rapidly with the system size because it requires the eigenvalues of the partial transposition of the density matrix. To avoid computational complexity, quantum mutual information is also frequently used. While it captures both quantum and classical correlations, mutual information is not an entanglement measure in the strict mathematical sense. 
Recently, the fermionic logarithmic negativity based on a partial time-reversal transformation has been suggested for fermion systems~\cite{Shapourian17a, Shapourian19a,Shapourian19b}. 
Since the partial time-reversal transformation of a fermionic Gaussian state remains Gaussian, the fermionic logarithmic negativity can be calculated directly from the $2L\times 2L$ matrix of generalized Green’s functions. 
Furthermore, it was shown to better capture the entanglement contributions from Majorana fermions.

On these grounds, in this article we mainly use the fermionic logarithmic negativity to analyze the mixed-state entanglement. To minimize the computational costs, we sometimes use the entanglement entropy for the average entanglement because quantum states along a trajectory are pure states.\footnote{In the several cases that we have checked, the behaviors of the logarithmic negativity and entanglement entropy turn out to be almost identical.}
Most of the results have been obtained by using the symbolic quantum simulation framework \textsf{Q3} \cite{Q3}; see also Ref.~\cite{ChoiMS22a}

\subsection{Kitaev Chain of Majorana Fermions}
\label{Paper::sec:KitaevChain}

\begin{figure}
\centering
\includegraphics[width=70mm]{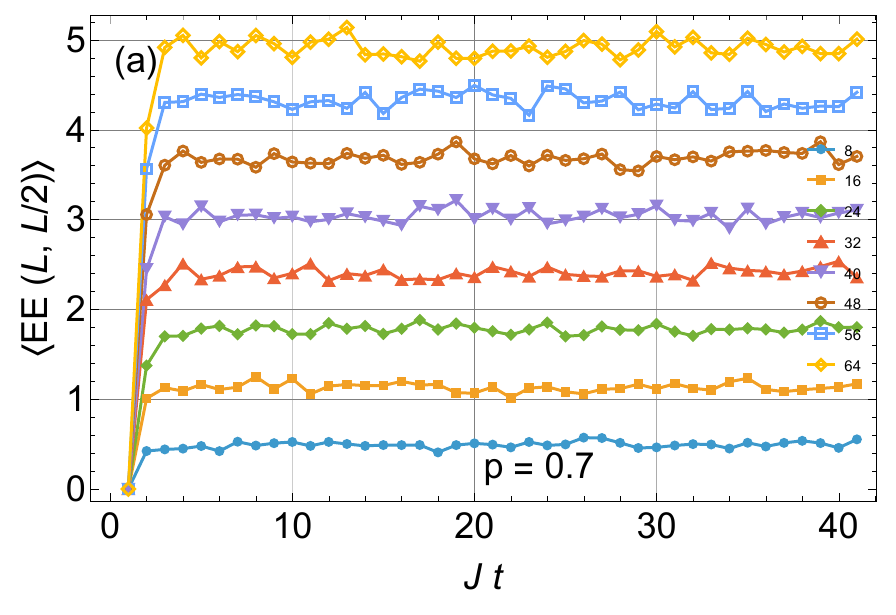} \quad
\includegraphics[width=70mm]{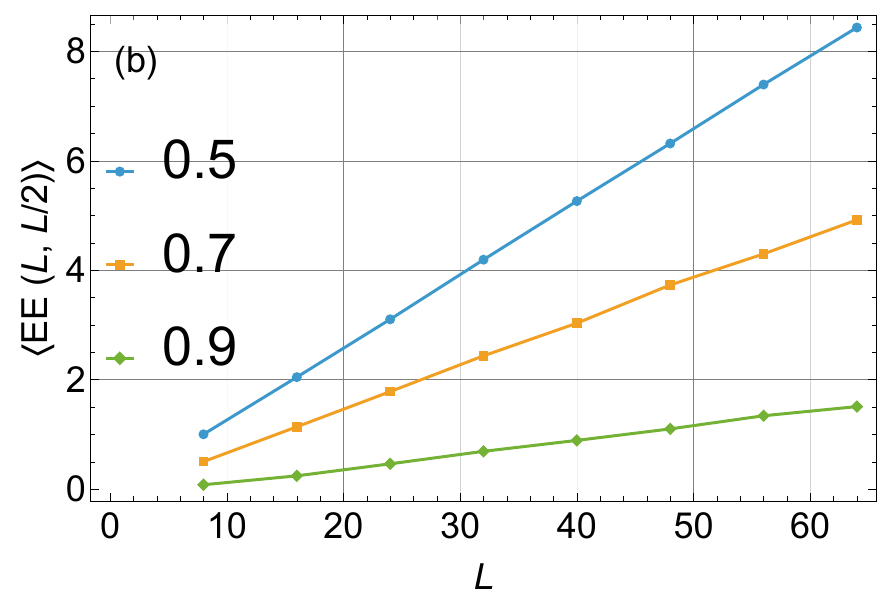}
\includegraphics[width=70mm]{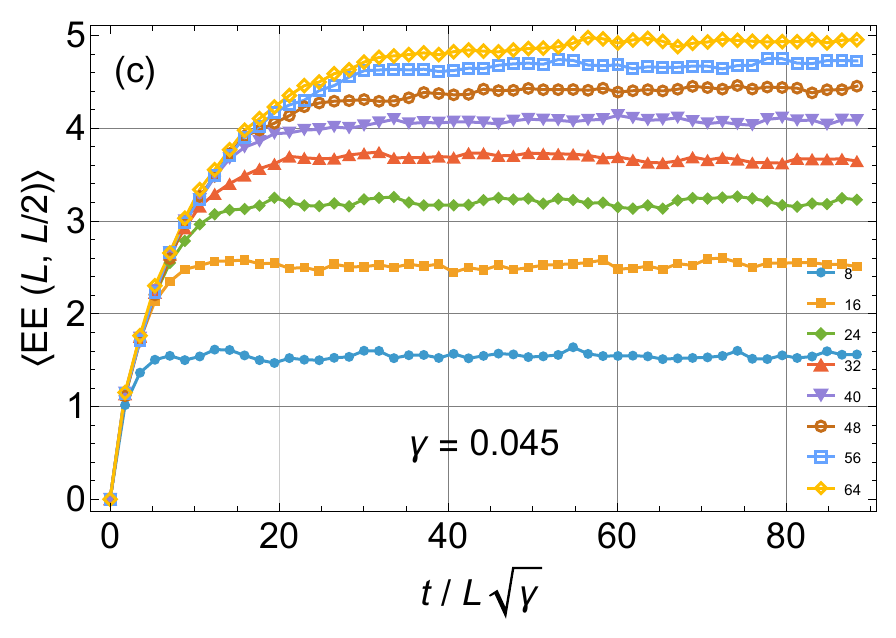} \quad
\includegraphics[width=70mm]{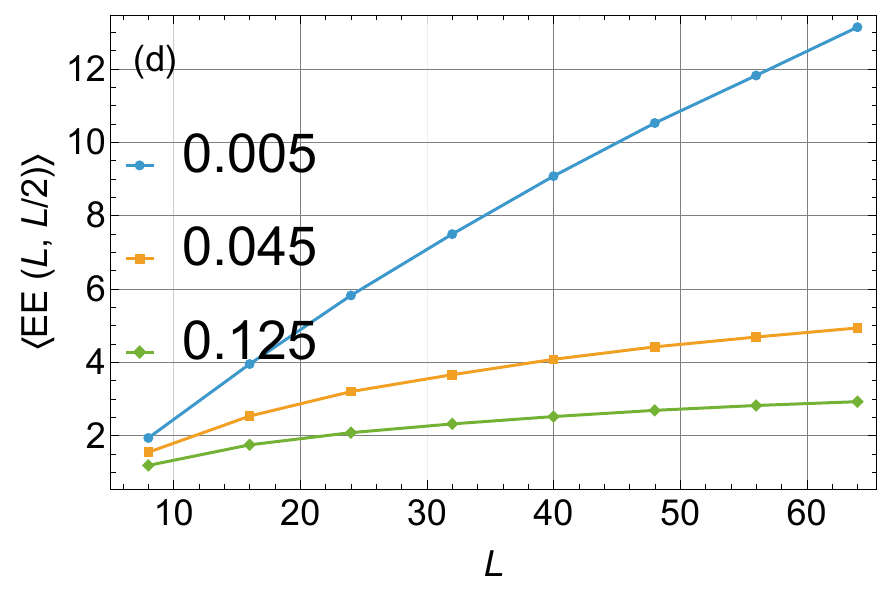}
\caption{The average entanglement entropy $\avg{\EE(L,L/2)}$ from the simulation of random circuits with the unitary layer governed by the Kitaev chain of Majorana fermions as a function of time (a, c) and size (b, d). 
In panels (a) and (b), the interaction time $\tau$ has been selected randomly from an exponential distribution over interval $[0,2\pi L/E_\mathrm{max}]$, where $E_\mathrm{max}\sim{J}$ refers to the largest energy scale of the system.
In panels (c) and (d), the interaction time $\tau$ has been selected randomly from an exponential distribution $P(\tau)\propto\exp(-L\gamma\tau)$.
The plot legends refer to system size $L$ in panels (a, c);
to measurement probability $p$ in panel (b); and
to $\gamma$ in panel (d).
Other parameter values are $J/\Delta=1$ and $\mu/\Delta=0.2$.}
\label{Paper::fig:4}
\end{figure}

\begin{figure}
\centering
\includegraphics[width=70mm]{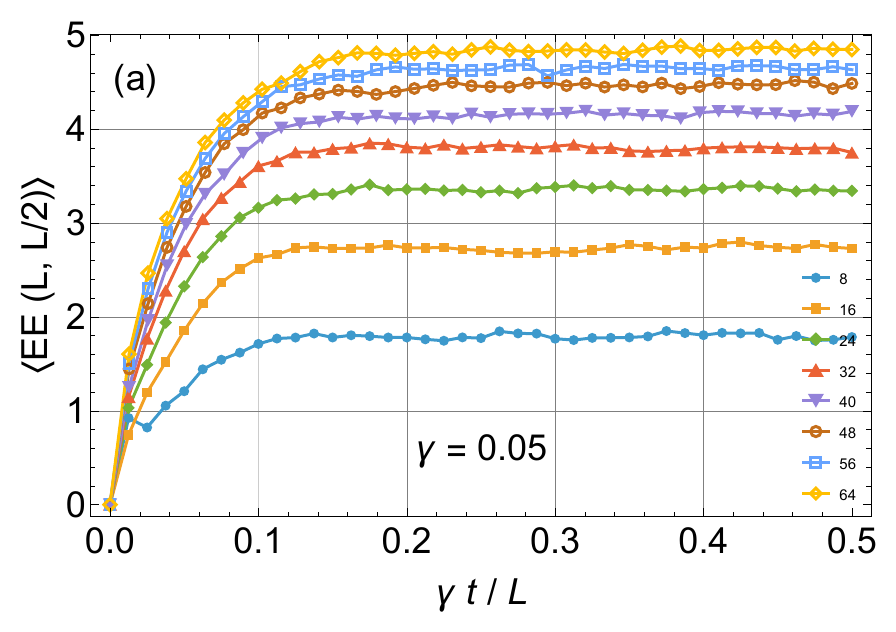}
\includegraphics[width=70mm]{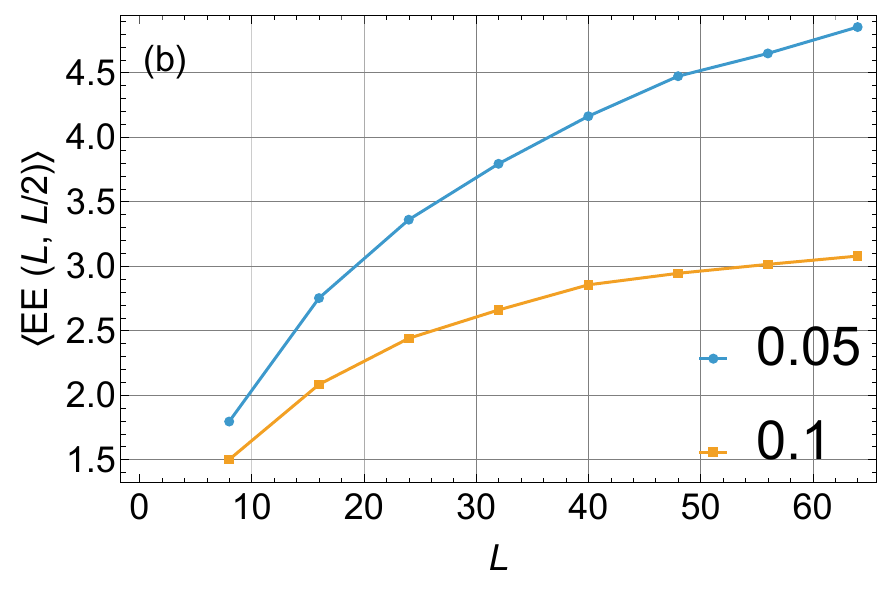}
\caption{The average entanglement entropy $\avg{\EE(L,L/2)}$ for the continuous monitoring of the Kitaev chain of Majorana fermions, calculated from the simulation based on the Lindblad equation, as a function of time (a) and size (b). The plot legends in panels (a) and (b) refer to the system size $L$ and monitoring rate $\gamma$, respectively. Other parameter values are $J/\Delta=1$ and $\mu/\Delta=0.2$.}
\label{Paper::fig:5}
\end{figure}

The Kitaev chain is a one-dimensional lattice model of spinless fermions featuring nearest-neighbor hopping and $p$-wave superconducting pairing \cite{Kitaev01a}. 
It has attracted much attention since it can host Majorana fermions bound to the two ends of the chain as topologically protected zero-energy edge modes in the topological phase. These Majorana fermions can be used for topological quantum computation \cite{Kitaev03a,Kitaev06b}.
The Kitaev chain is also frequently used to illustrate bulk-boundary correspondence, the existence of edge Majorana zero modes reflects the non-trivial topological classification of the bulk band structure \cite{Essin11a}.
The Kitaev chains have been experimentally realized, e.g., in semiconducting nanowire in proximity with conventional superconductors with external magnetic field along the nanowire \cite{Alicea12a,Mourik12a,Deng12a,Wilson12a}. Nevertheless, the direct evidence of Majorana fermions is still elusive.
In this work, we examine the quantum noisy dynamics of the Kitaev chain that is subject to interaction with the environment. To avoid redundancy, here we focus on the simulations of the fermionic random quantum circuits where the unitary layer is governed by the Hamiltonian of the Kitaev chain.
In the next subsection, we simulate the Lindblad equation with dissipative and projective Lindblad operators for the Hatano-Nelson model.

The Hamiltonian of the Kitaev chain of Majorana fermions is given by
\begin{equation}
\hatH_\mathrm{Kitaev} :=
-\mu\sum_{i=1}^L\hata_i^\dag\hata_i
-\frac{J}{2}\sum_{i=1}^{L-1}\left(\hata_i^\dag\hata_{i+1} + h.c.\right)
-\frac{\Delta}{2}\sum_{i=1}^{L-1}\left(\hata_i^\dag\hata_{i+1}^\dag + h.c.\right),
\end{equation}
where $\mu$ is the chemical potential, $J$ is the hopping amplitude, $\Delta$ is the superconducting $p$-wave pairing potential and $L$ is the system size.
The model exhibits a topological phase transition from the topological superconductor ($|\mu|<J$) to trivial superconductor ($|\mu|>J$).
The topological phase manifests itself with the Majorana fermion modes at the two ends of the chain.

We simulate fermionic random circuits (see Section~\ref{Paper::sec:RandomCircuits}) where the unitary layer is governed by the Hamiltonian of the Kitaev chain:
\begin{equation}
\label{Paper::eq:unitaryLayer}
\hatU(\tau) = \exp(-\ii\tau\hatH_\mathrm{Kitaev}),
\end{equation}
where $\tau$ is the interaction time to be chosen randomly and the local fermion modes are randomly selected for measurement in the measurement layer (we do not consider the fermion dissipation here).
For the random selection of interaction time $\tau$ and local fermion modes to be probed, we consider two different setups:
In the first setup, the \emph{interaction time} $\tau$ is selected randomly from the uniform distribution over the interval
\begin{math}
[0,2\pi L/E_\mathrm{max}],
\end{math}
while each local fermion modes are chosen with a fixed probability $p$.
The maximum interaction time has been set to 
\begin{math}
\tau_\mathrm{max}:=2\pi L/E_\mathrm{max},
\end{math}
where $E_\mathrm{max}$ refers to the largest energy scale of the system, so that all localized excitations are able to propagate from one end to the other.
As shown in Fig.~\ref{Paper::fig:4} (a) and (b), the entanglement entropy $\avg{\EE(L,L/2)}$ averaged over 200 trajectories of quantum states exhibits a surprisingly robust volume-law behavior even for measurement probability as large as $p=0.9$ although the absolute value decreases with $p$.
It seems that this strong entanglement is due to the sufficiently long interaction time $\tau$.

In the second setup, the interaction time $\tau$ is selected randomly from the exponential distribution
\begin{math}
P(\tau) \propto \exp(-L\gamma\tau)
\end{math}
and the measurement probability is fixed to $p=1/L$.
This setup is to emulate the physical situation of continuous monitoring. Since the average evolution time is given by $\avg{\tau}=1/L\gamma$, it is reasonable to assume that for each measurement layer, the probability $p$ for a fermion mode to be measured is given by $p=\gamma\avg{\tau}=1/L$.
Figure~\ref{Paper::fig:4} (c) and (d) show the average entanglement entropy $\avg{\EE(L,L/2)}$ in this setup. Unlike in the previous setup, the average entanglement entropy shows a sign of measurement induced phase transitions from the volume-law (for smaller $\gamma$) to area-law (for larger $\gamma$) behavior; see Fig.~\ref{Paper::fig:4} (d).
The investigation of the possibility and characteristic details of measurement-induced phase transitions is an interesting issue for future studies.

Finally, let us examine the problem of continuous monitoring of the Kitaev chain of Majorana fermions more rigorously. 
Within the Markov approximation, the situation is described by the Lindblad equation
\begin{equation}
\label{Paper::eq:KitaevMonitoring}
\frac{d\hat\rho}{dt} 
= -\ii\left[\hatH_\mathrm{Kitaev},\hat\rho\right]
+ \sum_i\gamma\left(
	\hata_i^\dag\hata_i\hat\rho\hata_i^\dag\hata_i
	+ \hata_i\hata_i^\dag\hat\rho\hata_i\hata_i^\dag
	- \hat\rho
\right),
\end{equation}
where $\gamma$ is the monitoring rate (or strength).
Note that Lindblad operators appear in pairs, $\hata_i^\dag\hata_i$ and $\hata_i\hata_i^\dag$, which add up to $\hatI$.
The effective non-Hermitian Hamiltonian is then given by
\begin{equation}
\hatH_\mathrm{non} := \hatH_\mathrm{Kitaev} - \ii\gamma L.
\end{equation}
That is, the non-Hermitian part is constant and leads to the uniform decay
\begin{math}
e^{-L\gamma{t}}
\end{math}
of the deterministic part of quantum state,
\begin{equation}
e^{-\ii{t}\hatH_\mathrm{non}}\hat\rho e^{-\ii{t}\hatH_\mathrm{non}^\dag}
= e^{-L\gamma{t}} e^{-\ii{t}\hatH_\mathrm{Kitaev}}\hat\rho
e^{+\ii{t}\hatH_\mathrm{Kitaev}}.
\end{equation}
Consequently, as argued above, one can expect that when discretized, the dynamics governed by \eqref{Paper::eq:KitaevMonitoring} is qualitatively the same as the random circuit of Kitaev chain with the interaction time $\tau$ randomly selected from the exponential distribution.
Figure~\ref{Paper::fig:5} shows the entanglement entropy $\avg{\EE(L,L/2)}$ averaged over the trajectories of quantum states from the simulation of the Lindblad equation for a continuous monitoring of fermion modes in the Kitaev model.
Comparing Fig.~\ref{Paper::fig:4} (c) and (d) with Fig.~\ref{Paper::fig:5} shows that they indeed agree qualitatively although there is no sign of a measurement-induced phase transition in the latter setup.
At this stage, we leave closer comparison and analysis of the two descriptions as an interesting open question for future studies.

In this subsection, we have focused on the average entanglement entropy and have not presented the mixed-state entanglement. This is because the latter is very small for the most of time in all cases and in particular, almost vanishes in the steady state.

\subsection{Hatano-Nelson Model}
\label{Paper::sec:HatanoNelson}

\begin{figure}
\centering
\includegraphics[width=70mm]{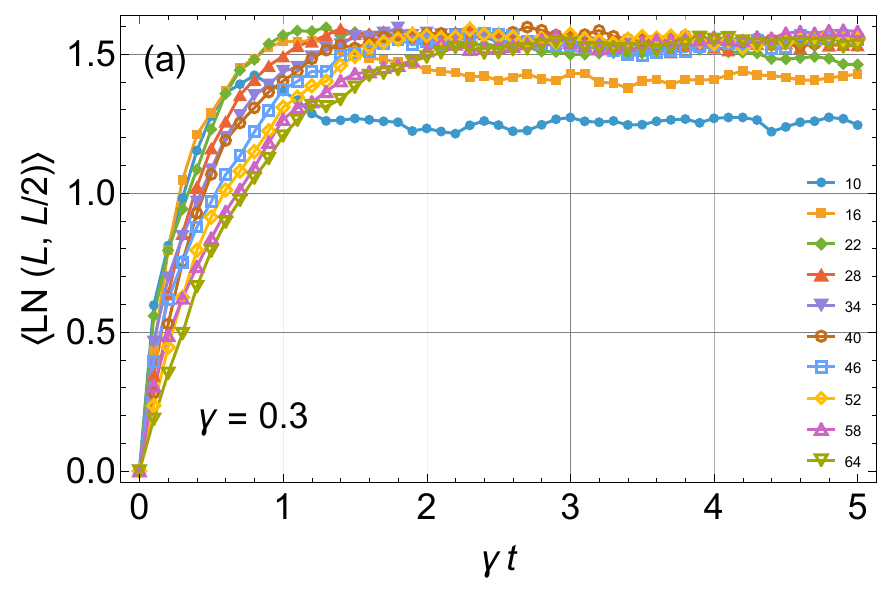}
\includegraphics[width=70mm]{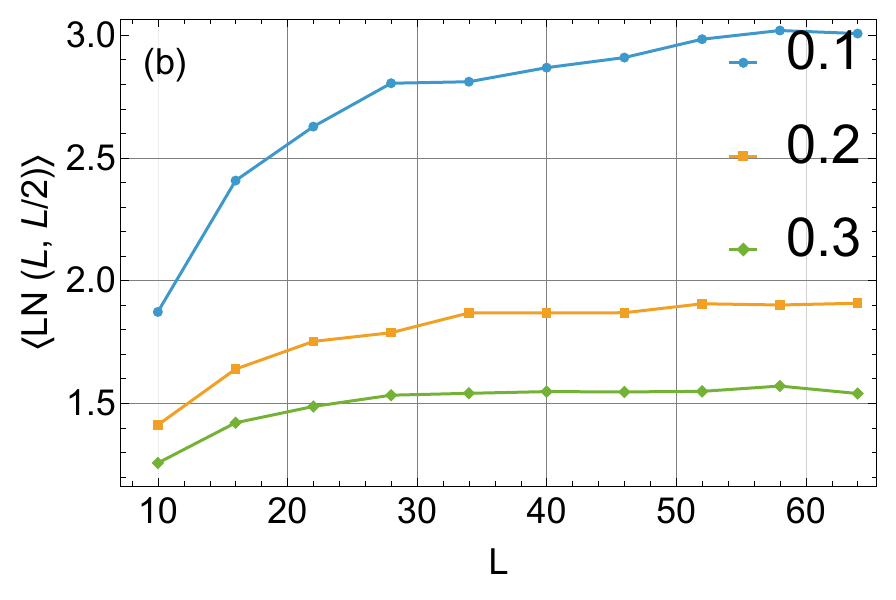}
\includegraphics[width=70mm]{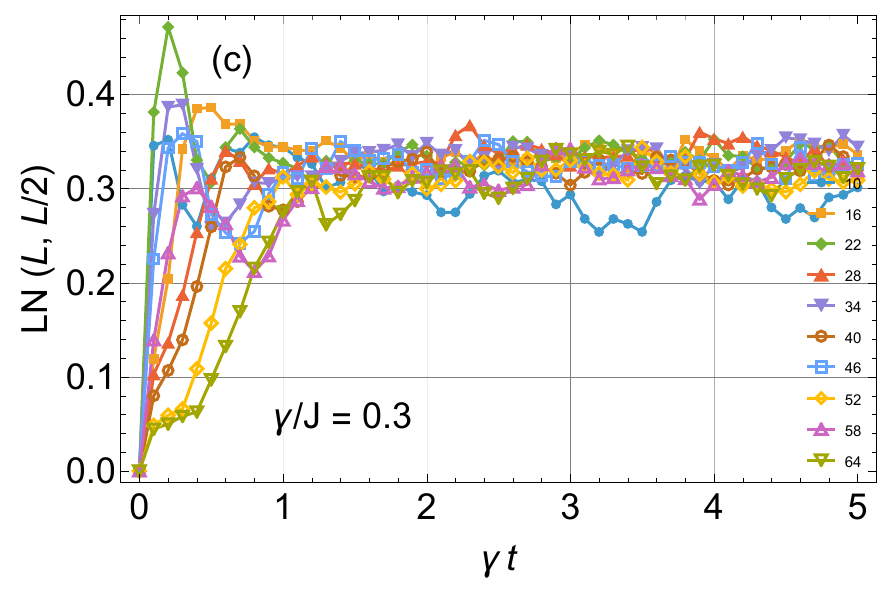}
\includegraphics[width=70mm]{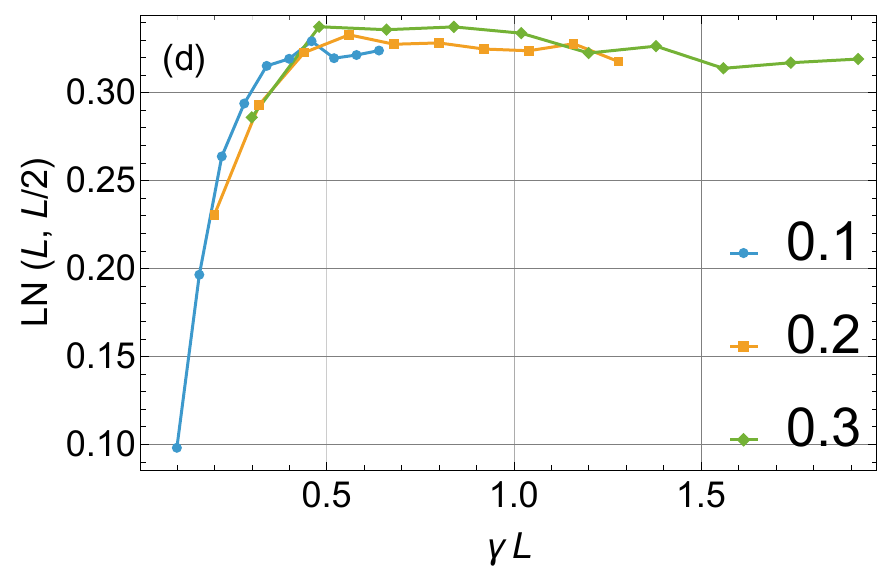}
\caption{Entanglement in the \emph{dissipative} Hatano-Nelson model with the dissipative realization: (a,b) The average logarithmic negativity $\avg{\LN(L,L/2)}$ as a function of time (a) and size (b).
(c,d) the logarithmic negativity $\LN(L,L/2)$ of the density matrix $\hat\rho(t)$ as a function of time (c) and size (d). In panels (a) and (c), the plot legends refer to $L$. In panels (b) and (d), the legends refer to $\gamma/J$.}
\label{Paper::fig:6}
\end{figure}

\begin{figure}
\centering
\includegraphics[width=70mm]{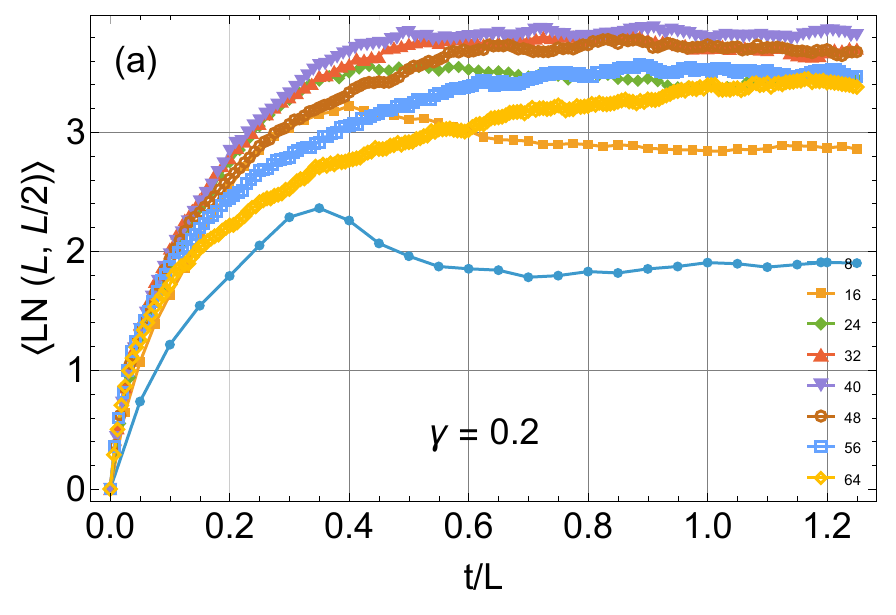}
\includegraphics[width=70mm]{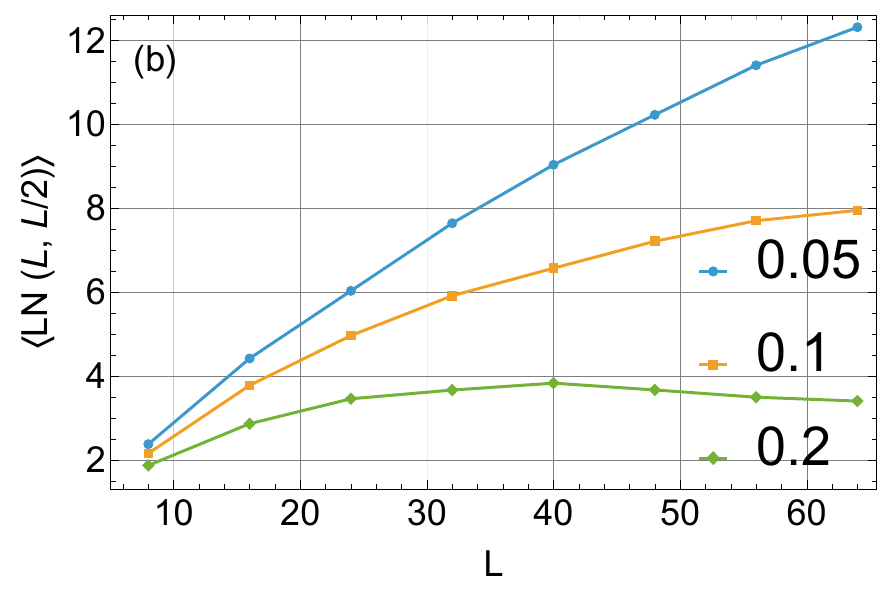}
\includegraphics[width=70mm]{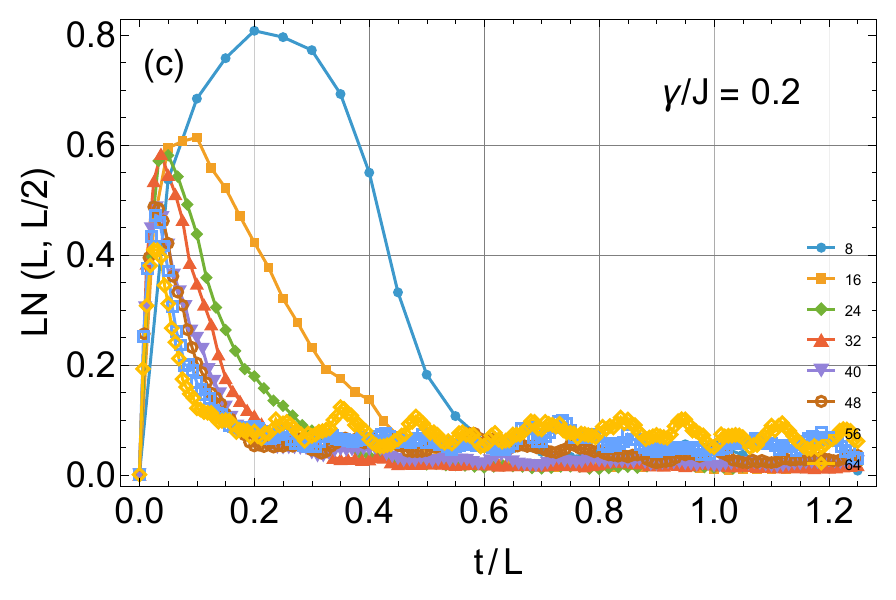}
\includegraphics[width=70mm]{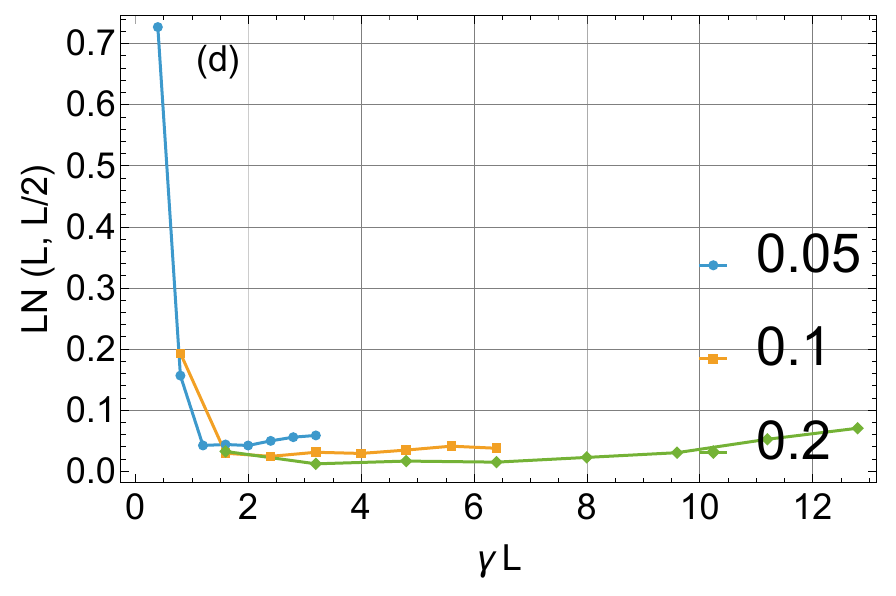}
\caption{Entanglement in the \emph{projective} Hatano-Nelson model with the dissipative realization: (a,b) The average logarithmic negativity $\avg{\LN(L,L/2)}$ as a function of time (a) and size (b).
(c,d) the logarithmic negativity $\LN(L,L/2)$ of the density matrix $\hat\rho(t)$ as a function of time (c) and size (d). In panels (a) and (c), the plot legends refer to $L$. In panels (b) and (d), the legends refer to $\gamma/J$.}
\label{Paper::fig:7}
\end{figure}

The Hatano-Nelson model is a paradigmatic model in the field of non-Hermitian physics, particularly in the context of non-Hermitian topology and the study of non-reciprocal systems. 
The model is characterized by its non-Hermitian Hamiltonian,
\begin{equation}
\label{Paper::eq:HatanaNelson}
\hatH_\mathrm{HN} = 
-\sum_{i=1}^{L-1}(J-\gamma)\hata_i^\dag\hata_{i+1}
-\sum_{i=1}^{L-1}(J+\gamma)\hata_{i+1}^\dag\hata_{i}
\end{equation}
for real constant $J$ and $\gamma$,
originally, that arises from introducing an imaginary vector potential into a metal with impurities\cite{Hatano96a}.

Recently, it was suggested to realize the Hatano-Nelson model as a result of interactions with the environment \cite{Kawabata23a}. To see the idea, consider the Lindblad equation 
\begin{equation}
\label{Paper::eq:HN:Lindblad}
\frac{d\hat\rho}{dt} = -i\left[\hatH,\hat\rho\right]
+ \sum_{k=-(L-1)}^{L-1} \left(\hatL_k\hat\rho\hatL_k^\dag
-\frac{1}{2}\hatL_k^\dag\hatL_k
\right),
\end{equation}
with the Hamiltonian
\begin{equation}
\hatH = -J\sum_{i=1}^{L-1}\left(\hata_i^\dag\hata_{i+1} + h.c.\right)
\end{equation}
and Lindblad operators
\begin{equation}
\label{Paper::eq:HN:dissipative}
\hatL_{\pm k} := \sqrt{\gamma}\,\hatb_{\pm k},
\end{equation}
where
\begin{equation}
\hatb_k := 
\frac{1}{\sqrt{2}}\left(\hata_k + \ii\hata_{k+1}\right)
\end{equation}
and
\begin{equation}
\hatb_{-k} := \frac{1}{\sqrt{2}}
\left(\hata_k^\dag + \ii\hata_{k+1}^\dag\right).
\end{equation}
Then, the effective non-Hermitian Hamiltonian
\begin{equation}
\hatH_\mathrm{non} := \hatH - \frac{\ii}{2}\sum_k\hatL_k^\dag\hatL_k 
= \hatH_\mathrm{HN}.
\end{equation}
takes the Hatano-Nelson model.
To study the non-Hermitian skin effects, Kawabata, Numasawa and Ryu \cite{Kawabata23a} examined the dynamics by taking into account only the Hamiltonian dynamics governed by the non-Hermitian Hamiltonian but ignoring the dissipative terms described by the Lindblad operators.
Experimentally, the dynamics in such a description can be observed only with the post-selection of the cases without any quantum jump process.

Naturally, there arise two issues:
First, the success probability for the desired post-selection is extremely small. Is the dynamics due to the pure Hatano-Nelson model physically feasible? 
Second, what are the effects of the quantum jump operators? In other words, does the entanglement phase transition observed in the non-unitary dynamics governed by the Hatano-Nelson model survive the processes described by the quantum jump operators?

Moreover, the way for the effective non-Hermitian Hamiltonian $\hatH_\mathrm{non}$ to lead to the Hatano-Nelson model is not unique. 
For example, the same Hatano-Nelson Hamiltonian is achieved with Lindblad operators of the projective type
\begin{equation}
\label{Paper::eq:HN:projective}
\hatL_{\pm k} := \sqrt{\gamma}\,\hatb_{\pm}^\dag\hatb_{\pm k}
\end{equation}
Although the effective non-Hermitian Hamiltonian is the same, the dynamics governed by the Lindblad equation~\eqref{Paper::eq:HN:Lindblad} with different Lindblad operators might be significantly different.
The effects of such difference in the Lindblad operators have not been clarified yet.

In this work, the Lindblad equation~\eqref{Paper::eq:HN:Lindblad} with Lindblad operators in Eq.~\eqref{Paper::eq:HN:dissipative} and that with Lindblad operators in Eq.~\eqref{Paper::eq:HN:projective} are called the \emph{dissipative} and \emph{projective} Hatano-Nelson models, respectively.
We have performed the numerical simulations and compare the entanglement dynamics in the two models.

Figure~\ref{Paper::fig:6} (a) and (b) show the logarithmic negativity $\avg{\LN(L,L/2)}$ averaged over 200 trajectories of quantum states as a function of time (circuit depth) and size, respectively, in the dissipative Hatano-Nelson model.
Figure~\ref{Paper::fig:6} (a) already compares sharply in one respect with the non-Hermitian Hamiltonian dynamics without the effects of Lindblad operators, which can be realized in principle by post-selection:
In the non-Hermitian Hamiltonian dynamics, the steady state is reached unusually late, at time $\gamma{t}\sim 100$ or even later. However, in our simulation taking into account the Lindblad operators,
the system reaches the steady state at time $\gamma{t}\sim 1$ as usually expected.

Figire~\ref{Paper::fig:6} (c) and (d) display the logarithmic negativity $\LN(L,L/2)$ of the density matrix $\hat\rho(t)$ constructed from the ensemble of pure states along the different trajectories.
Interestingly, in the Hatano-Nelson model, not only the average logarithmic negativity but also the mixed-state logarithmic negativity features a finite amount of entanglement. Furthermore, the mixed-state logarithmic negativity exhibits a universal scaling behavior as a function of re-scaled size $\gamma L$.

\begin{figure}
\centering
\includegraphics[width=80mm]{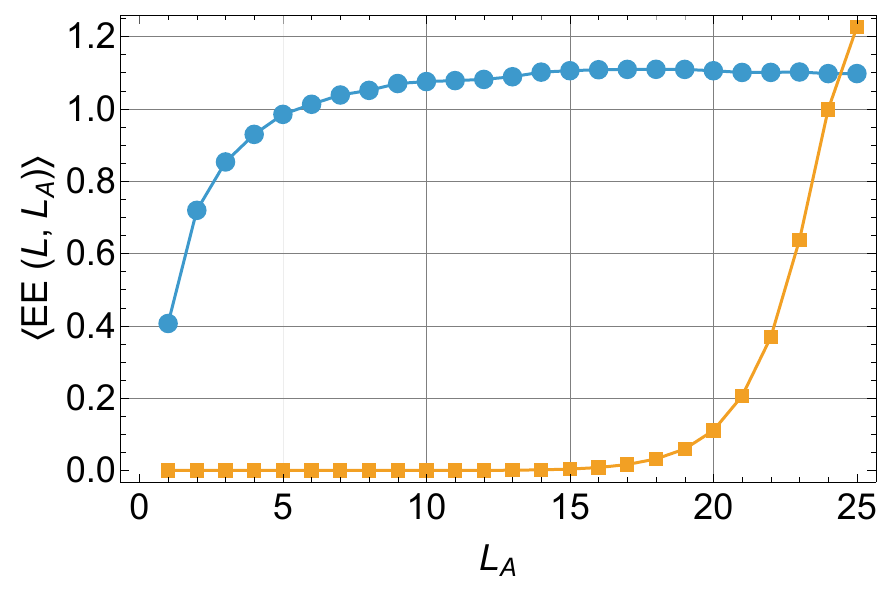}
\caption{
The average entanglement entropy $\avg{\EE(L,L_A)}$ (blue filled circles) as a function of subsystem size $L_A$ for fixed total system size $L=50$, averaged over 300 quantum trajectories from the Monte Carlo simulation of the Lindblad equation in the Hatano-Nelson model. For comparison, also shown is the entanglement entropy $\EE(L,L_A)$ (orange filled squares) from the (deterministic) dynamics governed by the non-Hermitian Hamiltonian. In both cases, non-Hermicity parameter is $\gamma=0.25J$.}
\label{Paper::fig:8}
\end{figure}

Figure~\ref{Paper::fig:7} shows the similar results as Fig.~\ref{Paper::fig:6} but for the projective Hatano-Nelson model.
The dynamical behaviors of two models are remarkably distinct:
First, the evolution of entanglement scales as $\gamma{t}$ in the dissipative model whereas it scales as $t/L$; compare Fig.~\ref{Paper::fig:6} (a) and (c) with Fig.~\ref{Paper::fig:7} (a) and (c).
Second, the averaged logarithmic negativity seems to exhibit a phase transition from the volume law (for smaller $\gamma$) to area law (for larger $\gamma$) in the projective model. In the dissipative model, it seems to remain in the area law.
Third, the mixed-state entanglement in the steady state casts even sharper difference. For the projective Hatano-Nelson model, the mixed-state entanglement is vanishingly small in the long-time limit [see Fig.~\ref{Paper::fig:7} (c) and (d)] while it is finite for the dissipative model.

Another interesting feature of the effects of the Lindblad operators is seen in the subsystem-size dependence of the average entanglement entropy $\avg{\EE(L,L_A)}$ for a fixed system size $L$ as shown in
Fig.~\ref{Paper::fig:8}.\footnote{Here, we examine the entanglement entropy to directly compare our results with those in \cite{Kawabata23a}. However, the logarithmic negativity also shows qualitatively similar behavior.}
When the Lindblad operators are taken into account, $\avg{\EE(L,L_A)}$ fast increases with $L_A$ from zero reaching the saturation value at relatively small $L_A$.
On the other hand, the average entanglement entropy $\avg{\EE(L,L_A)}$ from the (deterministic) non-Hermitian dynamics governed by the Hatano-Nelson Hamiltonian remains very small until $L_A\approx L/2$, where it sharply increases to the saturation value $\avg{\EE(L,L/2)}$.

\section{Conclusion}
\label{Paper::sec:Conclusion}

We have reviewed existing classical simulation methods for performing fermionic Gaussian operations and introduced new methods to address the gap by adhering to the fundamental theoretical framework established by Bravyi \cite{Bravyi05a} for the most general fermionic Gaussian processes. We have focused on the unified approach that can be applied to generic fermionic Gaussian operations, which is beneficial since the selection of simulation methods has often been based on an ad hoc choice, heavily influenced by the specific model and circumstances rather than a systematic approach.

The simulation framework discussed in this work is exclusively applicable to noninteracting (free) fermionic systems due to its inherent Gaussian structure. This structure relies on quadratic Hamiltonians and Wick’s theorem. Interacting systems, on the other hand, introduce higher-order correlations that surpass the accuracy of Gaussian approximations. Consequently, the method discussed in this work fails to provide exact simulability for interacting systems.
However, Gaussian approximations serve as benchmarks for weakly interacting models through perturbation theory. Ongoing extensions aim to disentangle circuits and impurity solvers. Furthermore, for weakly interacting fermions, perturbation theory expands around free Gaussian states. Interactions, such as density-density terms, are treated as corrections.
Indeed, Fermionic Gaussian circuits can disentangle interactions for tensor networks. This reduces bond dimensions in models like Anderson impurities. Superpositions of Gaussians or higher-order expansions quantify non-Gaussianity in weakly perturbed systems. These techniques aid in chaos detection and thermalization. Ground states of short-range interacting fermions often exhibit approximately Gaussian behavior in thermodynamic limits \cite{Dias24a,Matos21a,Wu25a}.

\section*{Acknowledgments}

Y.F. has been supported by the NSFC (Grants No. 12565004 and 12005011), Yunnan Province Foreign Expert Project (Grant No. 202505AP120020), and the Yunnan Fundamental Research Projects (Grant No. 202201AU070118).
H.C. and M.-S.C.\ have been supported by the National Research Function (NRF) of Korea (Grant Nos.~RS-2023-NR068116, 2023R1A2C1005588, RS-2024-00432563, RS-2024-00404854) and by the Ministry of Education through the BK21 Four program.
M.L. was supported by the NRF of Korea (Grant No. RS-2024-00354843).

\appendix
\section{Grassmann Variables}
\label{Paper::sec:Grassmann}

Grassmann variables are mathematical objects used primarily in theoretical physics and mathematics to describe systems with anti-commuting properties. Most notably, the behavior of fermions in quantum field theory can be completely described by Feynman path integrals over Grassmann variables.
A brief summary of Grassmann variables in the context of Gaussian dynamics of fermions is given in \cite{Bravyi05a} whereas a complete theory is described in the original paper \cite{Soper78a} and text books \cite{DiFrancesco96a} and \cite{Zinn-Justin02a}. Here, we provide a brief summary for self-containment.

\subsection{Grassmann Algebra}

Grassmann algebra $\calG_n$ of degree $n$ over the complex numbers $\BfC$ is constructed from a set of $n$ generators $x_i$ satisfying
\begin{equation}
\label{Paper::eq:cmm:Grassmann}
x_ix_j + x_jx_i =0.
\end{equation}
Note that $x_i^2=0$ for all generators $x_i$.
The Grassmann algebra $\calG_n$ is consists of all polynomials of generators; i.e., the monimials 
\begin{math}
x_{i_1}x_{i_2}\cdots x_{i_k}
\end{math}
of degree $k$ and all linear combinations of them with complex coefficients.
$\calG_n$ is a graded algebra:
A polynomial $f(x)$ is said to be even (odd) if it involves only monomials of even (odd) degree.

\subsection{Grassmann Derivative and Integral}

As for the complex variables, the derivative with respect to a Grassmann variable is a central concept in the calculus of anti-commuting variables.
It exhibits rules that differ notably from ordinary differentiation due to nilpotency and anti-commutation.

For a single Grassmann variable $x$, the derivative $\partial/\partial{x}$ is defined by
\begin{equation}
\frac{\partial}{\partial x} 1 = 0 \,,\quad
\frac{\partial}{\partial x} x = 1 
\end{equation}
with linearity implied.
For a set of Grassmann variables, the partial derivatives $\partial/\partial x_i$ are similarly defined by
\begin{equation}
\frac{\partial}{\partial x_i} 1 = 0 \,,\quad
\frac{\partial}{\partial x_i} x_j = \delta_{ij}
\end{equation}
as well as the Leibniz rule
\begin{equation}
\frac{\partial}{\partial x_i} [x_jf(x)] = 
\delta_{ij}f(x) - x_j\frac{\partial}{\partial x_i}f(x).
\end{equation}
The partial derivatives anti-commute with each other
\begin{equation}
\label{dd=0}
\drv{x_i}\drv{x_j} + \drv{x_j}\drv{x_i}=0.
\end{equation}

For the Grassmann variables, the derivative and integration are equivalent:
\begin{equation}
\int{dx_i} \equiv \frac{\partial}{\partial x_i}.
\end{equation}
For a multiple integration over Grassmann variables, we denote
\begin{equation}
\label{Bravyi05aX::eq:intMeasure}
\int\varD[x] := \int dx_1dx_2\cdots dx_{2n-1}dx_{2n}.
\end{equation}
Note that 
\begin{equation}
\int\varD[x]\;x_{2n}\cdots x_2x_1 = 1
\end{equation}
while
\begin{equation}
\int\varD[x]\;x_1x_2\cdots x_{2n} = (-1)^n.
\end{equation}

Under a change of variables
\begin{equation}
\label{Jacobian0}
y_i = \sum_{j=1}^n U_{ij} x_j,
\end{equation}
the derivative and integration measure change as
\begin{equation}
\label{Jacobian}
\drv{y_i} = \sum_{b=1}^{2n} U_{ji}^{-1} \drv{x_j},
\end{equation}
and
\begin{equation}
\label{Jacobian'}
\varD[y] = \left( \det{U}\right)^{-1} \, \varD[x],
\end{equation}
respectively.

A particularly useful Grassmann integrations are the following Gaussian integrations over Grassmann variables:
\begin{equation}
\label{Bravyi05aX::eq:GaussianIntegral1}
\int\varD[x]\,\exp\left(\frac{\ii}{2} \sum_{i,j}x_iA_{ij}x_j\right) 
= \Pf(-iA) = (-i)^n\Pf(A)
\end{equation}
and
\begin{equation}
\label{Bravyi05aX::eq:GaussianIntegral2}
\int\varD[x]\,\exp\left(\frac{\ii}{2}\sum_{i,j} x_iA_{ij}x_j + ix_iy_i\right) 
= (-i)^n\Pf(A)\exp\left\{+\frac{\ii}{2}\sum_{i,j}y_iA^{-1}_{ij}y_j]\right\},
\end{equation}
where $\Pf(A)$ denotes the Pfaffian of an $2n\times 2n$ anti-symmetric complex matrix $A$.
These Gaussian integrals are frequently used throughout this article:

\subsection[Grassmann Delta Function]{Grassmann $\delta$-Function}

By noting that
\begin{equation}
\int{dx}\,(x-y)f(x) = f(y),
\end{equation}
one can define the $\delta$-function
\begin{equation}
\label{Bravyi05aX::eq:delta}
\delta(x-y) := x - y
\end{equation}
for Grassmann variables $x$ and $y$.
Moreover, by expanding the exponential, one can get the integral representation of the Grassmann $\delta$-function
\begin{equation}
\label{Bravyi05aX::eq:deltaInt}
\delta(x-y) = \int{dz}\, e^{z(x-y)}
= \frac{1}{i}\int{dz}\,e^{iz(x-y)} ,
\end{equation}
which is similar to the usual $\delta$-function for complex variables.
Recall from Eq.~\eqref{Jacobian'} that
\begin{equation}
d(iz) = i^{-1}dz \quad (\neq i\,dz).
\end{equation}

\section{Noisy Quantum Channels}
\label{Paper::sec:NoisyChannels}

Here, we consider a general fermionic Gaussian channels described in the Kraus representation
\begin{equation}
\label{Paper::eq:Kraus}
\varF(\hat\rho) = \sum_\mu \gamma_\mu\hatF_\mu\hat\rho\hatF_\mu^\dag
\end{equation}
for some constants $\gamma_\mu>0$.
In this work, we consider the Kraus operators of either dissipative type
\begin{equation}
\label{Paper::eq:dissipative}
\hatF_\mu = \hatb_\mu
\end{equation}
or projective type
\begin{equation}
\label{Paper::eq:projective}
\hatF_\mu = \hatb_\mu^\dag\hatb_\mu
\end{equation}
for a dressed fermion mode
\begin{equation}
\hatb_\mu := \sum_{i=1}^{2n}w_{\mu,i}\hatc_i.
\end{equation}
Recall that each $\hatb_\mu$ is a valid Dirac fermion operator [see also Section~\ref{Paper::sec:FermiMeasurement}]
\begin{equation}
\{\hatb_\mu,\hatb_\mu^\dag\} = 1 \,,\quad
\{\hatb_\mu,\hatb_\mu\} = \{\hatb_\mu^\dag,\hatb_\mu^\dag\} = 0,
\end{equation}
which implies that [see also Eqs.~\eqref{Paper::eq:dressdedMode1a} and \eqref{Paper::eq:dressdedMode1b}]
\begin{equation}
u_{\mu}^Tv_{\mu} = 0 ,\quad
\left\|u_\mu\right\| = \left\|v_\mu\right\| = \frac{1}{2},
\end{equation}
where $u_{\mu}:=\re w_{\mu}$ and $v_{\mu}=\im w_{\mu}$, and 
\begin{math}
w_\mu^T := (w_{\mu,1},w_{\mu,2},\cdots,w_{\mu,2n}).
\end{math}
For distinct $\hatb_\mu$ and $\hatb_\nu$, however, there is no definite relation required between them; they refer to physically independent processes.
A physical decoherence process cannot increase the trace, and physically required is
\begin{equation}
\sum_\mu\gamma_\mu\hatF_\mu^\dag\hatF_\mu \leq 1.
\end{equation}
However, here we do not require this condition so that this method can be used as a part of larger numerical simulations.

Each term in the Kraus representation in Eq.~\eqref{Paper::eq:Kraus} describes an incoherent process concurring with \emph{relative} probability
\begin{math}
p_\mu 
= \gamma_\mu\Tr\left[\hatF_\mu\hat\rho\,\hatF_\mu^\dag\right]
= \gamma_\mu\Tr\left[\hat\rho\,\hatb_\mu^\dag\hatb_\mu\right]
\end{math}
for both the dissipative and projective case.
When the process $\hatF_\mu$ occurs, the quantum state is reduced to
\begin{equation}
\hat\rho \mapsto \varF_\mu(\hat\rho) 
:= \hatF_\mu\hat\rho\,\hatF_\mu^\dag.
\end{equation}
In the dissipative case, $\varF_\mu$ has been analyzed in Section~\ref{Paper::sec:FermiDissipation}. In the projective case, on the other hand, $\varF_\mu$ corresponds to $\varM_{b_\mu,1}$ in Section~\ref{Paper::sec:FermiMeasurement}.

In the above, we have analyzed the quantum state and probability assuming that a particular random process $\varF_\mu$ has occurred. Physically, this corresponds to generalized selective measurement \cite{Breuer02a}. This analysis is useful for the \emph{quantum trajectory approach} \cite{Plenio98a} as well; to be discussed in Section~\ref{Paper::sec:QME}.

In many cases, one may merely want to examine some average quantities. In such cases, it is sufficient to take the statistical mixture of each $\varF_\mu$ with corresponding relative probabilities $p_\mu$. In other words, we investigate the whole superoperator $\varF$. This is done simply by multiplying both sides of Eq.~\eqref{Paper::eq:Kraus} with $\hatc_i\hatc_j$ and then taking the traces while applying the Wick theorem \cite{Mahan00a}.
In the dissipative case, we get the correlation matrix $\varF(C)$ of the post-process state $\varF(\hat\rho)$
\begin{equation}
\varF(C) = P\left\{(CMC)^T_{ij} - (CMC)_{ij}
+ C\Tr\left[M C\right]\right\},
\end{equation}
where $P$ is the normalization factor of $\hat\rho$ and $M$ is a $2n\times 2n$ Hermitian matrix defined by 
\begin{math}
M := \sum_\mu \gamma_\mu M_\mu
\end{math} 
with
\begin{math}
M_\mu := w_\mu w_\mu^\dag .
\end{math}
In the projective case, on the other hand,
we get
\begin{multline}
\varF(C)
= 2P\sum_\mu\gamma_\mu^2
\left\{
\Tr[M_\mu C]C 
+ C^T\left(M_\mu - M_\mu^T\right)C
\right\}
\left(M_{\mu}-M_\mu^T\right)
\\ {}\quad
- 2P\sum_\mu\gamma_\mu^2\left(M_{\mu}-M_\mu^T\right)
\left\{
\Tr[M_\mu C]C
+ C^T\left(M_\mu - M_\mu^T\right)C
\right\}
\\ {}\quad
+ P\sum_\mu\gamma_\mu^2\left\{
\Tr[M_\mu C]C
+ C^T\left(M_\mu - M_\mu^T\right)C
\right\}.
\end{multline}

%%%% References
% \bibliographystyle{apsrev}
% \bibliography{References}
\bibliographystyle{apsrev}
\bibliography{physey}

\end{document}